\documentclass[
reprint,
twocolumn,
amsmath,amssymb,
aps,pra,notitlepage]{revtex4-1}
\usepackage{graphicx}
\usepackage{dcolumn}
\usepackage{bm}
\usepackage{braket}
\usepackage{xcolor}

\begin{document}
\title{Measurement based 2-qubit unitary gates for pairs of Nitrogen-Vacancy centers in diamond}

\author{Chenxu Liu}
\affiliation{Department of Physics and Astronomy, University of Pittsburgh, Pittsburgh, PA, 15260, USA}

\author{M. V. Gurudev Dutt}
\affiliation{Department of Physics and Astronomy, University of Pittsburgh, Pittsburgh, PA, 15260, USA}
\affiliation{Pittsburgh Quantum Institute, University of Pittsburgh, Pittsburgh, PA, 15260, USA}

\author{David Pekker}
\affiliation{Department of Physics and Astronomy, University of Pittsburgh, Pittsburgh, PA, 15260, USA}
\affiliation{Pittsburgh Quantum Institute, University of Pittsburgh, Pittsburgh, PA, 15260, USA}

\date{\today}

\begin{abstract}
The implementation of a high-fidelity two-qubit quantum logic gate remains an outstanding challenge for isolated solid-state qubits such as Nitrogen-Vacancy (NV) centers in diamond. In this work, we show that by driving pairs of NV centers to undergo photon scattering processes that flip their qubit state simultaneously, we can achieve a unitary two-qubit gate conditioned upon a single photon detection event. Further, by exploiting quantum interference between the optical transitions of the NV centers electronic states, we realize the existence of two special drive frequencies: a ``magic'' point where the spin-preserving elastic scattering rates are suppressed, and a ``balanced'' point where the state-flipping scattering rates are equal. We analyzed four different gate operation schemes that utilize these two special drive frequencies, and various combinations of polarization in the drive and collection paths. Our theoretical and numerical calculations show that the gate fidelity can be as high as 97\%. The proposed unitary gate, combined with available single qubit unitary operations, forms a universal gate set for quantum computing.
\end{abstract}

\maketitle

\section{Introduction} \label{sec:intro}
Quantum computers are expected to achieve considerable speedup as compared to classical computers~\cite{Grover1996,Shor1994,Shor1997,Simon1997}. A specific set of examples of this speedup ranges from polynomial for Grover's search algorithm, to sub-exponential for Shor's factorization algorithm, to exponential for Simon's algorithm. There has been significant progress on understanding complexity classes of quantum computation and their relation to the complexity classes of classical computation, see Ref.~\cite{Montanaro2016,QAZ} for a review of this progress. The key resource that enables quantum speedup is quantum entanglement. In order to generate and harness this resource, it is essential to build high fidelity multi-qubit quantum gates. 

The electronic spin associated with the nitrogen-vacancy (NV) centers in diamond is a promising qubit candidate for solid-state quantum computing. The spin states are well defined, have long spin relaxation and coherence times, and can be optically addressed for qubit initialization and readout for quantum operations. The qubits can be manipulated using either optical or microwave drive fields. However, a key missing ingredient for NV center quantum computing is an experimental demonstration of a high-fidelity 2-qubit unitary gate between NV centers at remote locations in the diamond lattice. 

There are two main directions that have been investigated for coupling pairs of NV centers. The first direction, which has been proposed theoretically~\cite{Yao2011}, relies on collective dynamics of spin-chains to deterministically generate couplings between two remote NV centers. The second direction, which has been investigated both theoretically and experimentally, generates entanglement between two NV centers using a measurement based method. Cabrillo et al. showed that measurement can be used to project two-qubit quantum state of atoms into an entangled state in Ref.~\cite{Cabrillo1999}. The idea of measurement-based entanglement generation was also theoretically proposed and studied in Refs.~\cite{Bose1999, Duan2001, Feng2003, Barrett2005, Zou2005, Lim2005}.  The quantum entanglement of two NV centers using measurement based method has also been explored experimentally. Bernien et al. observed quantum entanglement of spins of two NV centers~\cite{Bernien2013}. In a related work, Lee et al. demonstrated the entanglement of vibrational modes of two macroscopic diamonds (but not NV centers)~\cite{Lee2011}. Pfaff et al. experimentally entangled spin states of two NV-centers, which they used for quantum teleportation~\cite{Pfaff2014}. Hensen et al. experimentally performed the Bell inequality test via entangling two separated NV-center spin states~\cite{Hensen2015}. It is important to point out that the measurement of the photon in Refs.~\cite{Bernien2013,Pfaff2014,Hensen2015} is effectively a parity projector that projects the NV centers into a maximally entangled state. The limitation of this approach is that while it can be used to generate entanglement, it cannot be used to construct a 2-qubit unitary gate. 

Inspiration for our work comes from a previous theoretical proposal for constructing a measurement-based 2-qubit unitary gate using generic atoms~\cite{Protsenko2002}. Specifically, Protsenko et al. showed that quantum interference can be used to construct a 2-qubit unitary gate by controlling the relative phase of the photons emitted by the two atoms. This interference principle was later proposed for building 2-qubit gates between a pair of atoms in optical cavities coupled by linear optics~\cite{Zou2005}. 
 
In this paper, we propose an alternative measurement-based 2-qubit unitary gate for Nitrogen-Vacancy (NV) centers in diamond heralded by a single scattered photon. Further, we predict that there exists a ``magic'' frequency which suppresses spin-state preserving scattering transitions in favor of spin-flipping scattering transitions and a ``balance'' point where two spin-state flipping scattering transitions are equal. Utilizing these frequencies, in combination with a single mode diamond waveguide to collect and interfere the scattered photons, enables the proposed 2-qubit gate to achieve high fidelity and high success probability. For success probability approaching unity, the gate fidelity is $\sim 92\%$, while for fidelity approaching unity the success rate approaches $\sim 34\%$. 

A key advantage of our scheme is that, unlike the schemes in Refs.~\cite{Bernien2013, Pfaff2014,Hensen2015} that rely on two-photon Hong-Ou-Mandel interference, the success of our entangling unitary gate is heralded by a single photon detection. For example, if our protocol were implemented with bulk optics and microfabricated solid-immersion lenses in diamond as has been previously demonstrated, the detection probability is $p \sim 10^{-4}$~\cite{Bernien2013}, and with a conservative repetition rate $\sim 20$~kHz, this would result in a successful entangling gate operation every 0.5 seconds. By contrast, entanglement events occur every 10 minutes in the two-photon heralded schemes, which represents orders of magnitude improvement in the clock rate. With further improvements in collection efficiency using e.g. the nanobeam waveguides that we propose and analyze in this paper, and fast electronics, we can potentially achieve kHz - MHz clock rates that would be comparable to superconducting qubit quantum information processors.

This paper is organized as follows. In Sections~\ref{sec:setup} and~\ref{sec:transitions} we describe the main ingredients of our 2-NV center unitary gate. In Section~\ref{sec:setup}, we focus on the proposed experimental setup and how to use interference to construct a unitary gate. In Section~\ref{sec:transitions}, we argue for the existence of a ``magic'' frequency at which qubit state-preserving transitions are suppressed and a ``balance'' frequency at which qubit state-flipping transitions are balanced. We propose four gate operation schemes, three utilizing the ``magic'' frequency and one the ``balance'' frequency, and analyzed their fidelity, success probability and unitarity. In Section~\ref{sec:fidelity}, we analyze the success probability and fidelity of the 2-qubit unitary gate with possible experimental imperfections. We first build a qualitative understanding of the processes involved in the qubit dynamics and their effects on gate fidelity. Then we perform a quantitative analysis using the quantum trajectory method. We draw conclusions and present an outlook in section~\ref{sec:summary}. Details of the proposed waveguide geometry, photon collection efficiency, transition rate calculations and further discussion of gate fidelity can be found in the Appendix.

\section{Proposed experimental setup for a 2-NV unitary gate} \label{sec:setup}
In this section, we propose a realization of the scheme of Ref.~\cite{Protsenko2002} adapted for  NV centers. The 2-qubit unitary gate that was proposed in Ref.~\cite{Protsenko2002} has two main ingredients: (1) qubit state-flipping transitions that result in the emission of identical heralding photons, and (2) optical path-lengths from the qubits to the detector that differ by a $\pi/2$ phase difference. Ingredient (1) ensures that no matter the initial state of a qubit, whenever it absorbs a drive-photon and flips, it emits the desired heralding scattered photon. Ingredient (2) ensures that the measurement of a scattered photon corresponds to a unitary operation as opposed to a projection (e.g. ingredient 2 ensures that disentangled initial states map onto four distinct Bell states).

\begin{figure}[htbp]
\includegraphics[width=3.4 in]{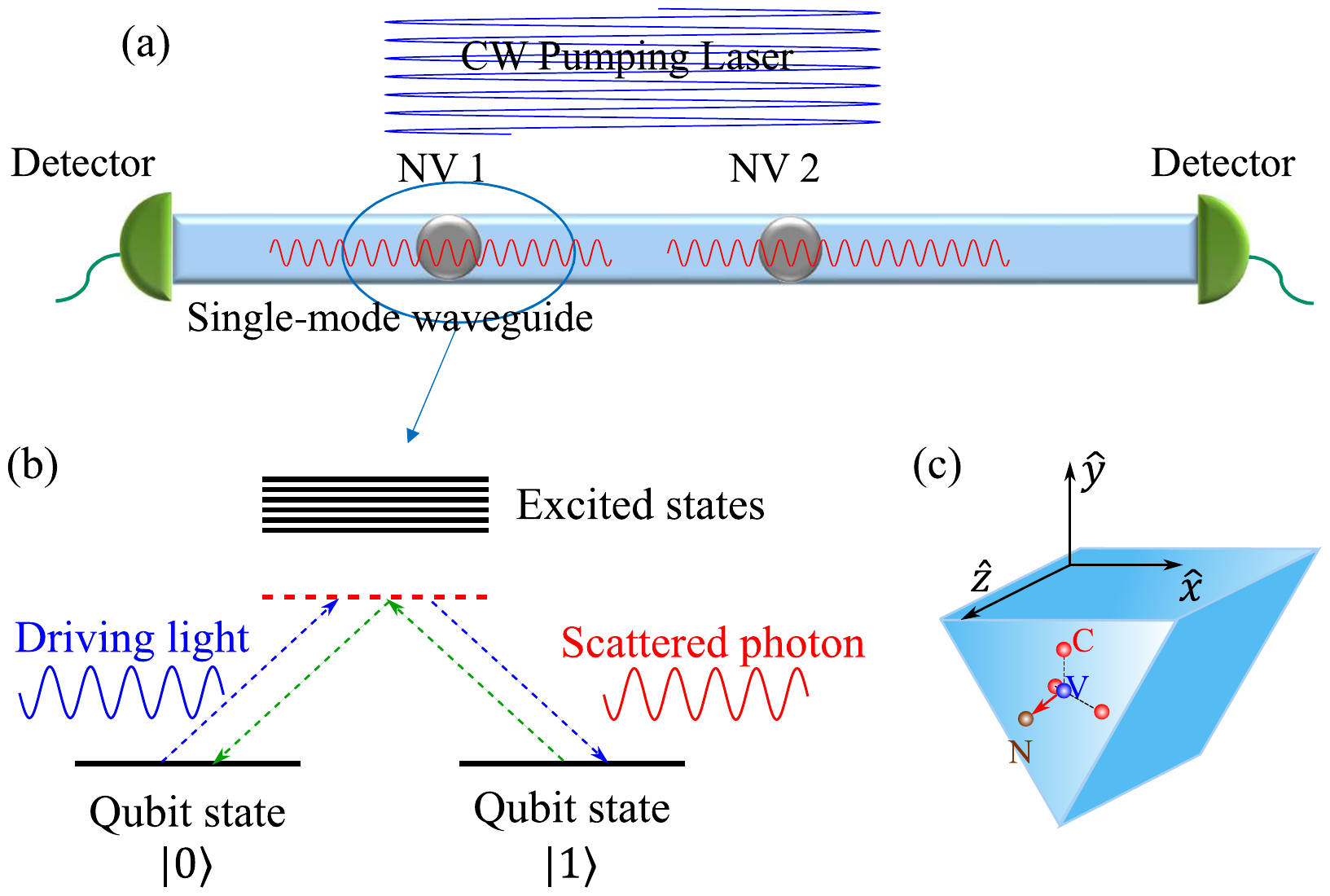}
\caption{(a) Schematic illustration for the proposed measurement-based two NV center quantum gate. A sketch of the level diagram of NV centers is shown in (b). The NV centers can undergo scattering transitions to flip the qubit states and emit scattered photons when they are driven by an off-resonance continuous wave (CW) pump laser. The two NV centers with quarter wavelength separation are in a single-mode diamond waveguide. The waveguide collects and interferes the scattered photons emitted from the NV centers. The detectors monitor the scattered photons collected by the diamond waveguide. The unitary gate operation is heralded by the detection of a photon. (c) The coordinate system of an NV center (red spheres -- carbon; blue -- vacancy; brown -- nitrogen), relative to the crystallographic axes of the diamond waveguide. The $\hat{x}$, $\hat{y}$ and $\hat{z}$ directions of the NV center match those of the waveguide, e.g. the $[\bar{1}\bar{1}\bar{1}]$ direction of diamond crystal (the red vector from the nitrogen to the vacancy) coincides with the axial direction of the waveguide.
}
\label{fig:MBEphonon}
\end{figure}

The experimental setup that we propose for a 2-qubit unitary gate using spin states of two NV centers is shown in Fig.~\ref{fig:MBEphonon}(a). The two NV centers are embedded into a single-mode diamond waveguide, and are selected so that they are separated by $(2n+1)/4$-wavelengths, where $n$ is an integer. The separation ensures that the emitted photons have $\pi/2$ phase difference when they are captured by the detectors. Both NV centers are aligned so that their $x$, $y$, and $z$-directions~\cite{Doherty2011} (i.e. the $[11\bar{2}]$, $[1\bar{1}0]$, $[\bar{1}\bar{1}\bar{1}]$,  direction of the diamond crystal) match the $x$, $y$, and $z$-directions of the waveguide (see Fig.~\ref{fig:MBEphonon}(c)). State-flipping transitions in both NV centers are pumped by a continuous-wave laser applied transverse to the waveguide [in Fig.~\ref{fig:MBEphonon}(a)]. The diamond waveguide collects and interferes the state-flipping scattered photons from the NV centers. Two detectors detect the photons collected by the waveguide from both ends to improve the detection efficiency. We note that depending on whether the detector on the left or on the right captures the photon we obtain slightly different unitary gates, which we discuss below.

We begin by reviewing why the $\pi/2$ phase is critical to achieve a unitary gate~\cite{Protsenko2002}. Assume that the NV centers have suitable state-flipping transitions which flip the qubit state between $\ket{0}$ and $\ket{1}$ and emit indistinguishable photons [Fig.~\ref{fig:MBEphonon}(b)]. Next, suppose that there is a phase difference of $\chi$ in the optical path from the two NV centers to the detector (on the right). Consider the two initial states $\ket{0,0}$ and $\ket{1,1}$. If the detector on the right clicks, the output states are $\ket{0,1}+e^{i \chi} \ket{1,0}$ and $\ket{1,0}+e^{i \chi} \ket{0,1}$. In order for our 2-qubit gate to be unitary, these two output states must be orthogonal, hence $\chi=\pi/2+n\pi$ where $n$ is an integer. Similar logic applies to the cases in which the initial states are $\ket{1,0}$ and $\ket{0,1}$. 

When the right detector clicks, the unitary 2-qubit gate is described by the matrix:
\begin{equation}
G_{r} = \frac{1}{\sqrt{2}} \left( \begin{array}{cccc}
0 & 1 & i & 0\\
1 & 0 & 0 & i\\
i & 0 & 0 & 1\\
0 & i & 1 & 0\\
\end{array}\right),
\label{eq:right_detector_gate}
\end{equation}
in the $\ket{0,0}$, $\ket{0,1}$, $\ket{1,0}$ and $\ket{1,1}$ basis.
On the other hand if the left detector clicks we obtain the gate described by the matrix:
\begin{equation}
G_{l} = \frac{1}{\sqrt{2}} \left( \begin{array}{cccc}
0 & i & 1 & 0\\
i & 0 & 0 & 1\\
1 & 0 & 0 & i\\
0 & 1 & i & 0\\
\end{array}\right).
\label{eq:left_detector_gate}
\end{equation}
Note that if we wanted to obtain $G_{r}$, but the left detector clicks instead, we can apply the single-qubit operation $X_{1} \otimes X_{2}$ to both qubits to convert the gate operation in Eq.~\eqref{eq:left_detector_gate} to the gate operation in Eq.~\eqref{eq:right_detector_gate}. We note that $G_{r}$ can be expressed in terms of the control-Z (CZ) gate and single-qubit gates as 
\begin{equation}
G_{r} = \frac{1+i}{\sqrt{2}} (H \otimes H) \left(S^{-1} \otimes S \right) \textrm{CZ} (H \otimes H),
\end{equation}
where $H$ is the Hadamard gate and $S$ is the single-qubit $\pi/2$ phase gate
\begin{equation}
H=\frac{1}{\sqrt{2}} \left( \begin{array}{cc}
1 &  1 \\
1 &  -1 
\end{array}\right), \quad\quad
S = \left( \begin{array}{cc}
1 & 0 \\
0 & i
\end{array}\right),
\label{eq:HS_gates}
\end{equation}
and therefore, our two-qubit gate, in combination with the available NV single-qubit gates, forms a universal gate set.

\section{Scattering transitions of an NV center for unitary 2-qubit gates} \label{sec:transitions}

\begin{figure}[tbp]
\includegraphics[width = 3.2 in]{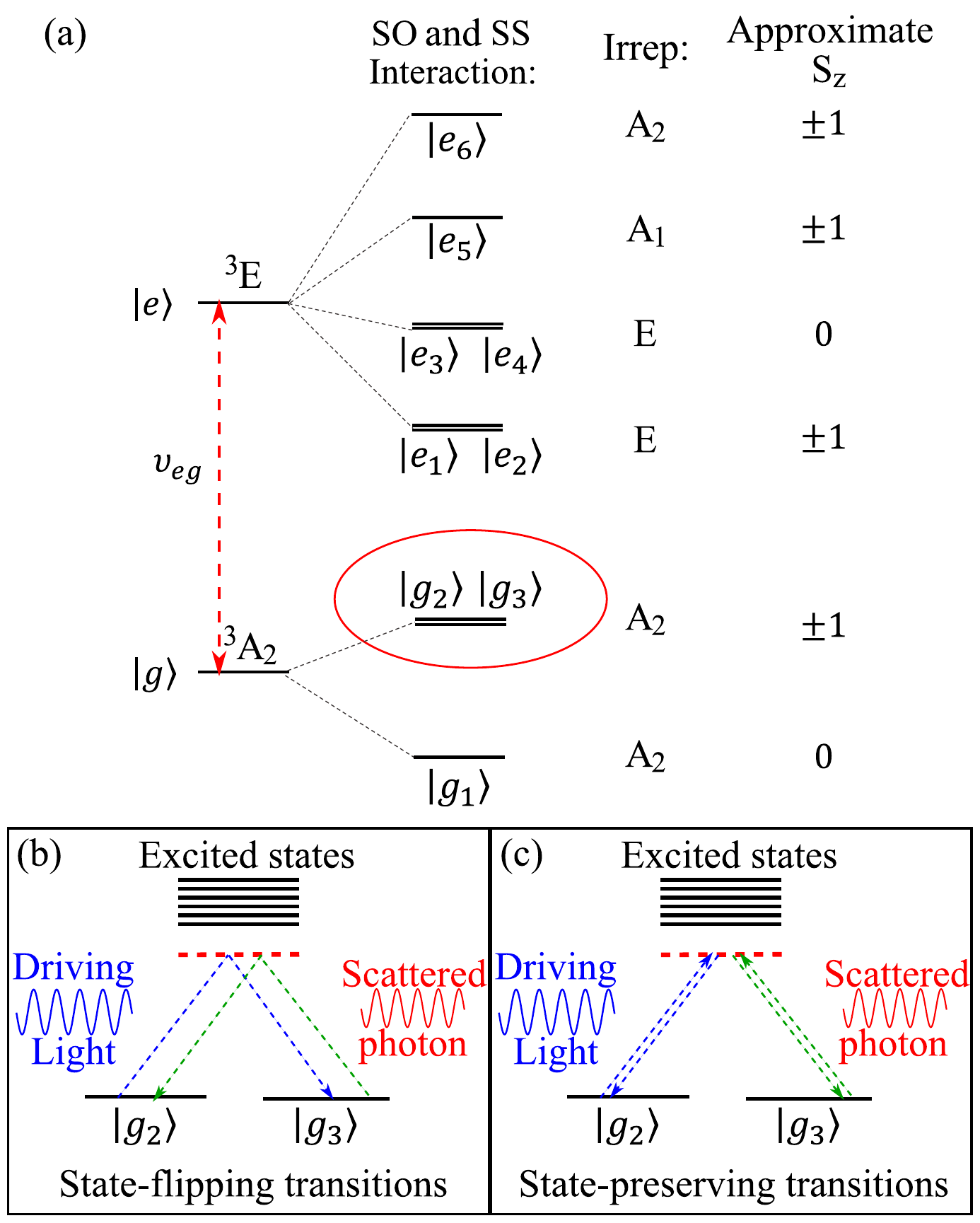}
\caption{The sketch of the level diagram of NV center electronic states is shown in (a). The electronic ground state $\ket{g}$ and excited state $\ket{e}$ splits due to the spin-orbit (SO) and spin-spin (SS) interactions with the corresponding irreducible representations (\textit{irrep}) of the $C_{3V}$ group and approximated $S_z$ quantum number. We choose to use state $\ket{g_2} = \frac{1}{\sqrt{2}} \left( \ket{+1} + \ket{-1} \right)$ and state $\ket{g_3} = \frac{i}{\sqrt{2}} \left( \ket{+1} - \ket{-1} \right)$ as the qubit states. We demonstrate the state-flipping transitions in (b) and state-preserving transitions in (c). The state-flipping transitions are the transitions that flips between qubit states $\ket{g_2}$ and $\ket{g_3}$. The other two transitions that does not flip NV states are the state-preserving transitions.}
\label{fig:levels}
\end{figure}

The main missing ingredient for constructing a 2-qubit gate with NV centers is finding suitable state-flipping transitions between qubit states of NV centers that emit indistinguishable scattering photons. In this section, we explore the electronic structure of NV centers and argue for the existence of suitable transitions. 

\begin{figure}[!htbp]
\includegraphics[width=3.4 in]{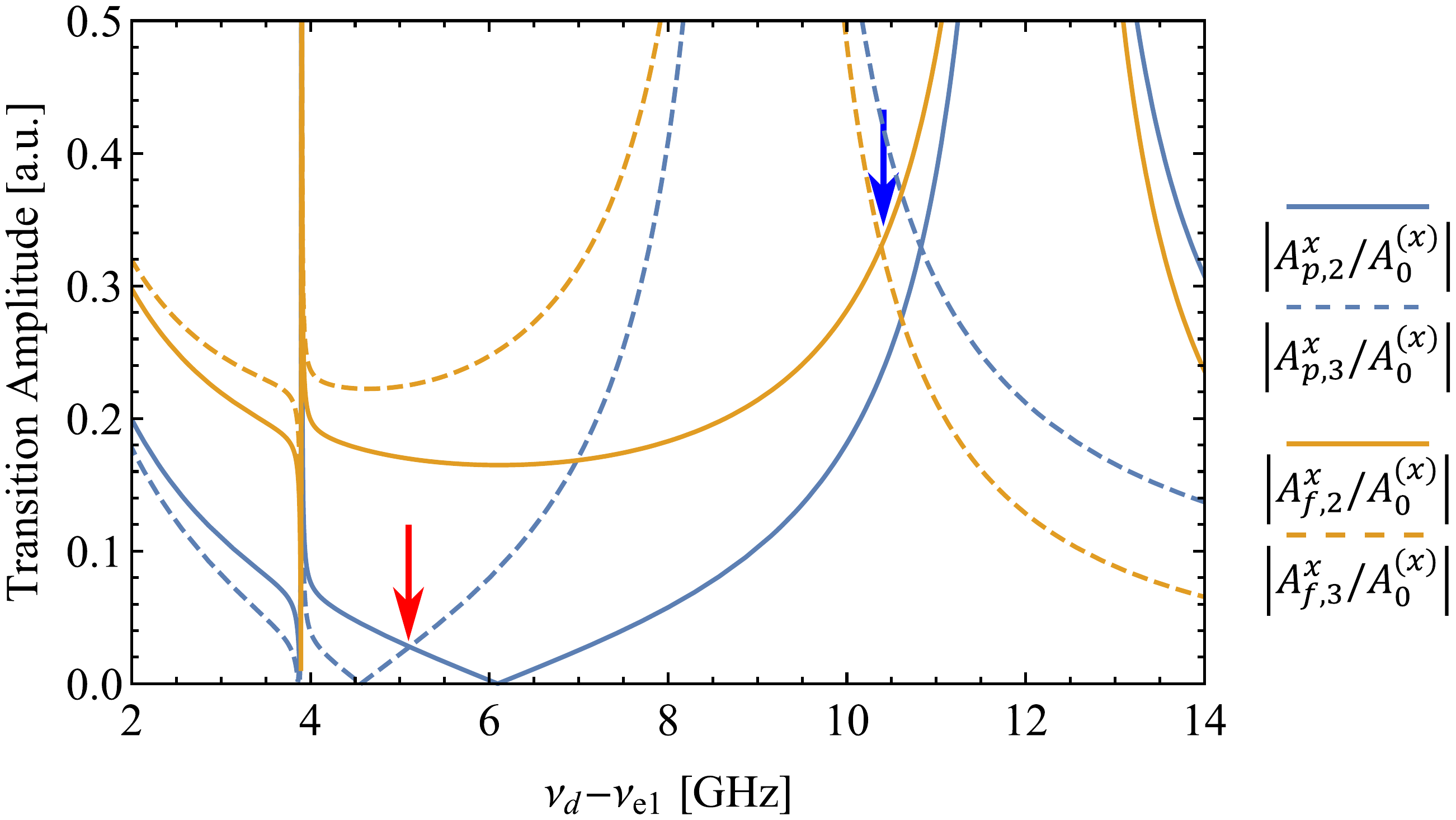}
\caption{The magnitudes of state-preserving and state-flipping transition amplitudes given in the R.H.S of Eq.~\eqref{eq:tran_amp_p} and Eq.~\eqref{eq:tran_amp_f} as we shift the driving light frequency $\nu_{d}$. There are two frequency points that draw our attention: (1) the ``magic'' point (labeled by red arrow) at which the two state-preserving transition amplitudes are strongly suppressed, (2) the ``balance'' point (labeled by blue arrow), where two state-flipping transition amplitudes are balanced.}
\label{fig:transition}
\end{figure}

Detailed information on electronic structures of NV centers can be found in Ref.~\cite{Doherty2011,Doherty2013} and in \ref{subsec:dipole} of our paper. The electronic levels, including fine-structure, of NV centers in diamond crystals without strain is shown in Fig.~\ref{fig:levels}(a). The electronic ground state of NV center is a spin triplet. The spin-spin interaction breaks the degeneracy of the NV electronic ground state and splits the state $\ket{g_1}=\ket{g,S_z=0}$ from the states $\ket{g_2}=\frac{1}{\sqrt 2} \left( \ket{g,S_z=+1}+ \ket{g,S_z=-1} \right)$ and $\ket{g_3}=\frac{1}{\sqrt 2} \left(\ket{g,S_z=+1}- \ket{g,S_z=-1} \right)$ by the zero field splitting $D/h = 2.87$~GHz. The manifold of excited states spans several GHz and consists of four discrete sets of states with six states in total [see Fig.~\ref{fig:levels}(a)]. These excited states can be labeled by the irreducible representation  of the $C_{3V}$ group and the $S_z$ quantum number. To simplify notation, we label them $\ket{e_j}$, where $j=1$ to $6$. We note that in the presence of spin-spin (SS) interactions $S_z$ is not a good quantum number for the lowest four excited states. However, as the SS interaction results in only a slight mixing between $S_z = \pm 1$ states and $S_z=0$ states we label the eigenstates $\ket{e_1}$, $\ket{e_2}$, $\ket{e_3}$ and $\ket{e_4}$ by the dominant $S_z$ component.

We propose to use the two-fold degenerate $S_z=\pm 1$ spin states, $\ket{g_2}$ and $\ket{g_3}$, as the logic $0$ and $1$ qubit states. We use scattering transitions pumped by an off-resonant laser to drive transitions between states $\ket{g_2}$ and $\ket{g_3}$ and hence flip the logic state [Fig.~\ref{fig:levels}(b)]. The scattered photons from the two state-flipping transitions have the same frequency because the states $\ket{g_2}$ and $\ket{g_3}$ are energetically degenerate. There are two more scattering transitions that can occur in principle, i.e. Rayleigh scatterings. These two transitions do not flip the qubit state [Fig.~\ref{fig:levels}(c)] and hence we call these transitions state-preserving transitions. The scattered photons emitted from these two transitions have the same frequency as the ones from state-flipping transitions. To ensure successful 2-qubit gate operation we must ensure that the detectors only click on state-flipping and not state-preserving transitions.

The two ingredients that go into the calculation of the optical transition rates are: (1) the dipole matrix elements between NV center ground and excited states and (2) the interference between virtual excitations of the various excited states. 

The results of the rate calculations for the state-flipping and state-preserving transitions, as a function of the drive frequency, are plotted in Fig.~\ref{fig:transition}. We find that as we tune the drive frequency the interference between virtual excitation paths results in the significant variation of the transition rates. We identify two special frequencies: first, there is a ``magic'' frequency for which the state-preserving transitions are approximately turned off. Second, there is a ``balance'' point frequency for which the two state-flipping transition rates are equal. We present the outline of the transition rate calculation in Section~\ref{subsec:outline_tran_rate} (the details are presented in~\ref{subsec:dipole}).  Next, we discuss four different schemes for building a 2-qubit gate using the two special drive frequencies and different configurations of polarizers in the collection path. Specifically, we discuss how different schemes can be used to optimize gate fidelity, success probability, and unitarity. In Sections~\ref{subsec:magic} and \ref{subsec:rotate} we discuss gates schemes $M1$, $M2$ and $M3$ that utilize ``magic'' frequency drive light. The three schemes differ by drive light polarization and collection path configuration which let us optimize either gate success probability or gate unitarity. In Section~\ref{subsec:balance_point}, we discuss the gate scheme $B1$, that utilizes driving light frequency which makes the two state-flipping transitions balanced. We summarize the configurations of the four gate operation schemes in Table.~\ref{tab:schemes}.

\begin{table}[!htbp]
\caption{\label{tab:schemes} The configurations of the four gate operation schemes. We list the driving light frequency and polarization, and the collection path polarizer orientation for each schemes. Polarizations that appear in brackets are alternative to the ones that appear with no brackets.}
\begin{ruledtabular}
\begin{tabular}{l c c c}
Gate    & Drive      & Drive        & Collection path \\
schemes & frequency  & polarization & filter polarization \\
\colrule
$M1$ & ``magic'' point    & $\hat{x} \, (\hat{y})$  & $\hat{y} \, (\hat{x})$ \\
$M2$ & ``magic'' point    & $\hat{x} + \hat{y} \, (\hat{x} - \hat{y})$ & $\hat{x} - \hat{y} \, (\hat{x} + \hat{y})$ \\
$M3$ & ``magic'' point    & $\hat{x} + \hat{y} \, (\hat{x} - \hat{y})$ & $\hat{x} + \hat{y} \, (\hat{x} - \hat{y})$ \\
$B1$ & ``balanced'' point & $\hat{x} \, (\hat{y})$  & $\hat{y} \, (\hat{x})$ \\
\end{tabular}
\end{ruledtabular}
\end{table}

\subsection{Transition rate calculation: interference of virtual excitation paths} 
\label{subsec:outline_tran_rate}
The dipole moment matrix after taking spin-orbital (SO) and spin-spin (SS) interaction into account can be written as
\begin{equation}
\frac{\hat{\mathbf{p}}}{p_0}=\left(
\begin{array}{cccccc}
-F_{21} \hat{x} & F_{21} \hat{y} & F_{22} \hat{x} & F_{22} \hat{y} & -F_{23} \hat{y} & F_{23} \hat{x} \\
F_{21} \hat{y} & F_{21} \hat{x} & -F_{22} \hat{y} & F_{22} \hat{x} & F_{23} \hat{x} & F_{23} \hat{y} 
\end{array}
\right).
\label{eq:dipole}
\end{equation}
Here, $p_0$ is the scale of the dipole moment; the matrix is written in the basis $\hat{p}_{ij} = \bra{g_i} \hat{\mathbf{p}} \ket{e_j}$, where $i=1$ for $\ket{g_2}$, $i=2$ for $\ket{g_3}$ and $j=1$ to $6$ for excited states $\ket{e_1}$ to $\ket{e_6}$; and the factors $F_{21}$, $F_{22}$, $F_{23}$ are three dimensionless parameters from the microscopic NV center Hamiltonian, $F_{21}=0.7062$, $F_{22} = 0.0363$, $F_{23}=1/\sqrt{2}$ (see Ref.~\cite{Doherty2011} and~\ref{subsec:dipole} for details).

The scattering transition rates between the states $\ket{g_2}$ and $\ket{g_3}$ can be calculated using second order Fermi's golden rule. According to Eq.~\eqref{eq:dipole}, if the driving light is linearly polarized along $\hat{x}$ or $\hat{y}$ direction, the photons from state-preserving transitions have the same polarization as the incoming light, while the photons from the state-flipping transitions have orthogonal polarization. Therefore, the state-flipping scattering photons can be distinguished from state-preserving scattering photons by polarization. In general, the result of perturbation theory can be expressed as 
\begin{equation}
\ket{g_j}\ket{\hat{\sigma}_1}_i \xrightarrow{H_{\textrm{scatter}}}
A_{p,j}^{\hat{\sigma}_1} \ket{g_j} \ket{\hat{\sigma}_1}_o + A_{f,j}^{\hat{\sigma}_1} \ket{g_k} \ket{\hat{\sigma}_2}_o
\label{eq:trans_amp}
\end{equation}
where $j,k=1,2$ and $j \neq k$, $A$'s represent the transition amplitudes, the incoming drive light is in the polarization state $\hat{\sigma}_1$, and the outgoing light in the waveguide is in the polarization state $\hat{\sigma_1}$ or $\hat{\sigma_2}$~\footnote{As we discuss in~\ref{subsec:waveguide}, the transverse directions of the NV centers are aligned to the transverse directions of the waveguide, which leads to the $\hat{x}$ and $\hat{y}$ directions of the dipole moment to couple to two different guided modes of the waveguide.}. 

Let us consider the case in which the driving light is linearly polarized along either $\hat{x}$ or $\hat{y}$ direction, and hence $\langle\hat{\sigma}_1|\hat{\sigma}_2\rangle=0$. We present the generic case in~\ref{subsec:transition_rates}. 
Assuming the driving light frequency is $\nu_d$, based on the dipole moment matrix, the state-preserving transition amplitudes can be worked out as,
\begin{equation}
\begin{aligned}
\frac{A^{x}_{p,2}}{{A}_{0}^{(x)}} = \frac{A^{y}_{p,3}}{A_{0}^{(y)}} & = \frac{1}{\Delta_1} F_{21}^2 + \frac{1}{\Delta_3} F_{22}^{2} + \frac{1}{\Delta_6} F_{23}^2   \\
\frac{A^{x}_{p,3}}{A_{0}^{(x)}} = \frac{A^{y}_{p,2}}{A_{0}^{(y)}}& = \frac{1}{\Delta_2} F_{21}^2 + \frac{1}{\Delta_4} F_{22}^{2} + \frac{1}{\Delta_5} F_{23}^2 
\end{aligned}
\label{eq:tran_amp_p}
\end{equation}
where the $\Delta_i = \epsilon_{e,i}-\epsilon_{g}-h \nu_{d}$ is the energy mismatch, $\epsilon_{e,i}$, $\epsilon_{g}$ are the energy of the excited state $\ket{e_i}$ and the ground state $\ket{g_2}$, $\ket{g_3}$. As we shift the driving light frequency $\nu_{d}$, the energy detuning of each excited level ($\Delta_i$) changes. Two scale factors, $A_{0}^{(x)}$ and $A_{0}^{(y)}$, are defined as $A_{0}^{(\sigma)} = p_{0}^2 E_{d,\sigma} \mathcal{E}_0 u_0$, where $E_{d,\sigma}$ is the driving light electric field along $\hat{\sigma}$ direction, $\mathcal{E}_0 = \sqrt{h \nu_{d}/(2 \varepsilon_0)}$ is the electric field associated with a single photon in the waveguide, $u_0$ is the normalized waveguide mode profile at the location of the NV centers (see Eq.~\eqref{eq:quantize_guide_mode} in~\ref{subsec:transition_rates}) We assume that the electric fields of the two guided modes have the same $u$ at the location of the NV centers. In~\ref{subsec:transition_rates}, we show that there is a region inside the waveguide where the two modes have balanced coupling to the NV centers. See~\ref{subsec:transition_rates} for details. In the following discussion, we assume these two parameters, $A_{0}^{(x)}$ and $A_{0}^{(x)}$, are equal. We also notice that the equality relations 
\begin{equation}
\frac{A^{x}_{p,2}}{A_{0}^{(x)}} = \frac{A^{y}_{p,3}}{A_{0}^{(y)}}, \quad \frac{A^{y}_{p,2}}{A_{0}^{(y)}} = \frac{A^{x}_{p,3}}{A_{0}^{(x)}}
\label{eq:pres_amp_equality}
\end{equation} 
hold if $\vert \bra{g_2} \hat{\mathbf{p}} \ket{e_i} \vert = \vert \bra{g_3} \hat{\mathbf{p}} \ket{e_i} \vert$ for all excited states. 

Similarly, the state-flipping transition amplitudes are,
\begin{equation}
\begin{aligned}
\frac{A^{x}_{f,2}}{A_{0}^{(x)}} = \frac{A^{y}_{f,3}}{A_{0}^{(y)}}& = -\frac{1}{\Delta_1} F_{21}^{2} - \frac{1}{\Delta_{3}} F_{22}^{2} +\frac{1}{\Delta_6} F_{23}^{2}  \\
\frac{A^{x}_{f,3}}{A_{0}^{(x)}} = \frac{A^{y}_{f,2}}{A_{0}^{(y)}}& = \frac{1}{\Delta_2} F_{21}^{2} + \frac{1}{\Delta_4} F_{22}^{2} - \frac{1}{\Delta_5} F_{23}^{2} 
\end{aligned}
\label{eq:tran_amp_f}
\end{equation}
Note that these two equality relations $\frac{A^{x}_{f,2}}{A_{0}^{(x)}} = \frac{A^{y}_{f,3}}{A_{0}^{(y)}}$ and $\frac{A^{y}_{f,2}}{A_{0}^{(y)}} = \frac{A^{x}_{f,3}}{A_{0}^{(x)}}$ do not rely on the special symmetry in dipole moment elements. We plot the magnitudes of the R.H.S. of the Eq.~\eqref{eq:tran_amp_p} and Eq.~\eqref{eq:tran_amp_f} in Fig.~\ref{fig:transition} as we shift the driving light frequency $\nu_d$.

\subsection{M1 2-qubit gate scheme: ``Magic'' frequency, $\hat{x}$-polarized drive light} 
\label{subsec:magic}
As we shift the driving light frequency $\nu_d$, we notice that there is a ``magic'' point where both state-preserving transition rates are highly suppressed because of the destructive interference between the virtual paths through the different excited states (see Fig.~\ref{fig:transition}). 

When we use an $\hat{x}$ polarized driving light, the scattered photons from state-preserving transitions are polarized along the $\hat{x}$ direction, while the polarization of the photons from state-flipping transitions are orthogonal, i.e. along $\hat{y}$. We can use a polarizer to further filter the state-flipping photons from the state-preserving photons. Heralding on the photons coming through the polarizer, we achieve a 2-qubit gate on the NV centers. This is our proposed gate scheme $M1$.

At the ``magic'' frequency the transition amplitudes satisfy $A_{p,2}^{x} = -A_{p,3}^{x} > 0$, $A_{f,2}^{x}<0$ and $A_{f,3}^{x} > 0$. Therefore we define $A_{p,2}^{x} = - A_{p,3}^{x} = A_{p} >0$ and define 
\begin{equation}
A_{1} \equiv \vert A_{f,2}^{x}\vert = -A_{f,2}^{x}, \, A_{2} \equiv \vert A_{f,3} \vert=A_{f,3}.
\label{eq:trans_amp_mgcpt}
\end{equation}
Since the state-preserving transition amplitudes satisfies $A_{p,2}^{x} = -A_{p,3}^{x}>0$, we can also define $A_p = A_{p,2}^{x} = -A_{p,3}^{x}$.

At the ``magic'' frequency, however, the two state-flipping transition amplitudes are not balanced. These two unbalanced transition amplitudes cause the resulting gate to be slightly non-unitary. Assuming the polarizer is perfect and the right detector captures the heralding photon, the 2-qubit gate is described by the matrix,
\begin{equation}
G^{(1),\text{ub}}_r= \left( \begin{array}{cccc}
0 & A_2 & i A_2 & 0 \\
-A_1  & 0 & 0 & i A_2  \\
-i A_1 & 0 & 0 & A_2 \\
0 & -i A_1 & -A_1 & 0 \\
\end{array}\right)
\label{eq:gate_1}
\end{equation}
in the basis $\ket{g_2;g_2}$, $\ket{g_2;g_3}$ and $\ket{g_3;g_2}$ and $\ket{g_3;g_3}$. If we have two balanced state-flipping transitions, i.e. $A_{1} = A_{2}$, after proper normalization, the gate operation is a 2-qubit unitary gate, and it can be written as
\begin{equation}
G^{(1),\text{b}}_r = \frac{1}{\sqrt{2}}\left( \begin{array}{cccc}
 0  & 1  & i & 0 \\
-1  & 0  & 0 & i\\
-i & 0  & 0 & 1 \\
 0  & -i  & -1 & 0 \\
\end{array}\right)
\label{eq:gate_matrix_ideal}
\end{equation}
where we write down the gate operation in the same basis as Eq.~\eqref{eq:gate_1}. Notice that this gate operation is different from the one shown in Eq.~\eqref{eq:right_detector_gate}. This is because of the negative state-flipping transition amplitude $A_{f,2}^{x}$. This gate is also equivalent to CZ gate combining with single qubit gates as, 
\begin{equation}
G^{(1),\text{b}}_r = \frac{1+i}{\sqrt{2}}\left((S^{-1} H) \otimes (S H)\right) \textrm{CZ} \left((H S^{-1}) \otimes (H S)\right)
\end{equation}
where $S$, $H$ are single qubit phase gate and Hadamard gate shown in Eq.~\eqref{eq:HS_gates}. 
When the two transition amplitudes are not balanced, i.e. $A_{1} \neq A_{2}$, the gate operation shown by Eq.~\eqref{eq:gate_1} is not unitary. 

we calculate the entanglement fidelity of our 2-qubit gate. Notice that both the entanglement fidelity and the average fidelity, which can be relatively easily calculated, it is proven to be related~\cite{Horodecki1999,Nielsen2002}. Here 
Here, we use the entanglement fidelity for the quantum channel to evaluate the quality of our gate~\cite{Nielsen2004}. Consider a quantum channel $\mathcal{E}$ acting on quantum system $Q$. Suppose there is another quantum system $R$ and there is a maximally entangled state $\ket{\phi}$ on system $QR$. The entanglement fidelity is defined as:
\begin{equation}
F_{e}(\mathcal{E}_Q) = \bra{\phi} \left[\mathcal{I}_{R}\otimes \mathcal{E}_{Q}\right](\ket{\phi}\bra{\phi}) \ket{\phi} 
\end{equation}
where $\mathcal{I}_{R}$ is the action of the identity operation on the system $R$ and $\mathcal{E}_Q$ is the action of the quantum channel on the system $Q$. In our scenario we considered a 2-qubit gate operation instead of a quantum channel to transfer a quantum state. We adapt the above definition to the entanglement fidelity of an imperfect quantum gate operation $\mathcal{G}$ as compared to the ideal quantum gate operation $\mathcal{U}$ via:
\begin{equation}
F_{e}(\mathcal{U}_Q,\mathcal{G}_Q) = \bra{\phi} \left[\mathcal{I}_{R}\otimes (\mathcal{U}^{\dagger}_{Q} \circ \mathcal{G}_{Q})\right](\ket{\phi}\bra{\phi}) \ket{\phi} 
\end{equation}
where $\mathcal{U}_{Q}$ is the desired unitary gate operation on system $Q$ and $\mathcal{G}_{Q}$ is the non-ideal gate operation, notation $\circ$ stands for composition of gate operations. Note that the quantum operation $\mathcal{G}$ should be trace preserving, though it may be non-unitary. For example, the quantum operation $\mathcal{G}$, corresponding to the gate $G^{(1),\textrm{ub}}_r$, on the system density operator $\rho$ is,
\begin{equation}
\mathcal{G}^{(1),\textrm{ub}}_r (\rho) = \frac{ G^{(1),\textrm{ub}}_r \rho \, \left[{G^{(1),\textrm{ub}}_r}\right]^{ \dagger} }{\textrm{Tr} \left[G^{(1),\textrm{ub}}_r \rho \, \left[{G^{(1),\textrm{ub}}_r}\right]^{\dagger} \right] } 
\end{equation}

To apply the definition above to a two-qubit system, we need another two-qubit system in order to construct a maximally entangled state over the four-qubits. We choose the state $\ket{\phi} = \sum_{j=1}^{4} \frac{1}{2} \ket{j_{R}}\ket{j_{Q}}$, where $\ket{j}$ is $\ket{g_2;g_2}$, $\ket{g_2;g_3}$, $\ket{g_3;g_2}$, $\ket{g_3;g_3}$ for $j=1$ to $4$ on corresponding 2-qubit systems. With the transition amplitudes calculated at the ``magic'' frequency as $A_1 \sim 0.1696$ and $A_2 \sim 0.2252$, the entanglement fidelity of our gate operation shown in Eq.~\eqref{eq:gate_1} is 
\begin{align}
F_e (\mathcal{G}^{(1),\textrm{b}}_r, \mathcal{G}^{(1),\textrm{ub}}_r) = \frac{(A_1+A_2)^2}{2(A_1^2+A_2^2)} \sim 0.981.
\end{align}

\subsection{M2 \& M3 2-qubit gate schemes: ``Magic'' frequency, $\hat{x}\pm\hat{y}$-polarized drive light} \label{subsec:rotate}
In this subsection, we discuss two schemes, $M2$ and $M3$, to perform the 2-qubit gate operation at the ``magic'' frequency. In the $M2$ scheme, we choose $(\hat{x}+\hat{y})$ polarized driving light with a $(\hat{x}-\hat{y})$ polarizer (mode filter) on the collection path. In the $M3$ scheme, we also choose $(\hat{x}+\hat{y})$ polarized diving light, but use $(\hat{x}+\hat{y})$ polarizer. Scheme $M2$ results in a slightly non-unitary gate with higher success probability as compared to scheme $M3$. Scheme $M3$, on the other hand, results in a 2-qubit gate that is exactly unitary, but has a low success probability. We note that similar schemes can be constructed with the alternative choice of $(\hat{x}-\hat{y})$ polarized drive light.

To understand the gate operation when we rotate the driving light polarization, we need to know the scattered photon polarization. Suppose the driving photon is in state $\ket{\hat{\sigma}_d} = \cos(\theta) \ket{\hat{x}}_i + \sin(\theta)e^{i \phi} \ket{\hat{y}}_i$. According to Eq.~\eqref{eq:trans_amp}, if an NV center is initialized in $\ket{g_2}$ state, the final states of the NV center and the scattered photon are 
\begin{equation}
\begin{aligned}
\ket{g_2} \otimes &\ket{\hat{\sigma}_d} \xrightarrow{H_{\textrm{scatter}}} \ket{\Psi_{2;\hat{\sigma}_d}} \\
& =\ket{g_2} \left( \cos(\theta) A^{x}_{p,2} \ket{\hat{x}} + \sin(\theta) e^{i \phi} A^{y}_{p,2} \ket{\hat{y}} \right) \\
& +\ket{g_3} \left( \cos(\theta) A^{x}_{f,2} \ket{\hat{x}} + \sin(\theta) e^{i \phi} A^{y}_{f,2} \ket{\hat{y}} \right)
\end{aligned}
\end{equation}
where we use notation $\ket{\Psi_{2,\hat{\sigma}_d}}$ to show the final state of the NV center and the scattered photon when the initial state of NV center is $\ket{g_2}$ and the drive light is $\ket{\hat{\sigma}_d}$. Using the $\hat{\sigma}_d$ polarized driving light to pump the transition from a single NV center in state $\ket{g_2}$, the state-preserving scattered photon is in state $\ket{\hat{\sigma}^{p}_{2}} \propto \cos(\theta) A^{x}_{p,2} \ket{\hat{x}} + \sin(\theta) e^{i \phi} A^{y}_{p,2} \ket{\hat{y}}$ up to a normalization constant, while the state-flipping scattered photon is in state $\ket{\hat{\sigma}^{f}_{3}} \propto \cos(\theta) A^{x}_{f,2} \ket{\hat{y}} + \sin(\theta) e^{i \phi} A^{y}_{f,2} \ket{\hat{x}}$. Similarly, the state of the photons from the scattering process with initial state $\ket{g_3}$ are $\ket{\hat{\sigma}^{p}_{3}} \propto A^{x}_{p,3} \cos(\theta)\ket{\hat{x}} + A^{y}_{p,3} \sin(\theta)\ket{\hat{y}}$ for state-preserving photons, and $\ket{\hat{\sigma}^{f}_{3}} \propto A^{x}_{f,3} \cos(\theta) \ket{\hat{y}} + A^{y}_{f,3} \sin(\theta) e^{i\phi} \ket{\hat{x}}$ for state-flipping photons.

As we rotate the driving light from the $\hat{x}$ to $\hat{y}$ direction, the scattered photons from two state-flipping transitions do not have the same polarization, i.e. $\langle\hat{\sigma}^f_2 \vert \hat{\sigma}^f_3\rangle \neq 1$ after the proper normalization of states $\ket{\hat{\sigma}^f_2}$ and $\ket{\hat{\sigma}^f_3}$. This occurs because the transition amplitudes $A^{x}_{f,2} = A^{y}_{f,3} \neq A^{y}_{f,2} = A^{x}_{f,3}$. Therefore, we need a polarizer on the collection path to erase the quantum information carried by the state-flipping photons. If the NV center in state $\ket{g_i}$ is pumped with $\ket{\hat{\sigma}_{d}}$ drive light and the polarizer in the collection path only allows photons in the state $\ket{p}=-\sin(\alpha) \ket{\hat{x}}_o + \cos(\alpha) e^{i \beta} \ket{\hat{y}}_o$, then the final state of the NV center heralded by a photon detection is $\ket{\psi^{\hat{p}}_{i,\hat{\sigma}_d}} \propto \langle p \ket{\Psi_{i;\hat{\sigma}_d}}$. 

\begin{figure}
\includegraphics[width=6cm]{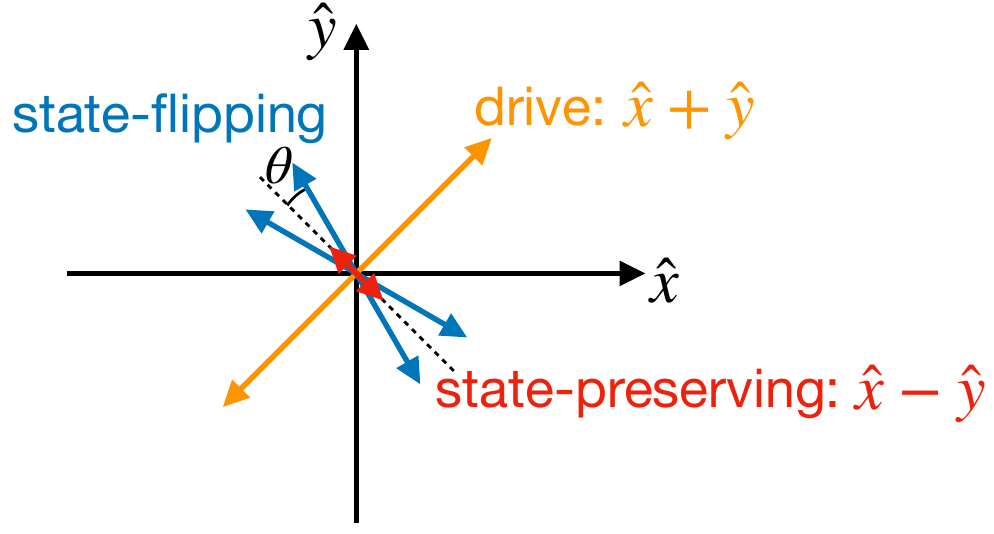}
\caption{Polarization diagram for drive light polarized along $(\hat{x}+\hat{y})$. State-preserving scattered photons are polarized along $(\hat{x}-\hat{y})$. State flipping scattered photons are polarized in a direction $\pm\theta$ away from $(\hat{x}-\hat{y})$ (the sign being determined by the initial state of the NV center). }
\label{fig:polarization}
\end{figure}

By rotating the driving light polarization to the direction $(\hat{x}+\hat{y})$, i.e. $\ket{\hat{\sigma}_d}_i = \ket{+} = \frac{1}{\sqrt{2}}(\ket{\hat{x}}_i+\ket{\hat{y}}_i)$,  we balance the state-flipping transition rates. In this case, the state-preserving photons are polarized along $(\hat{x}-\hat{y})$ direction, and the state-flipping photons are polarized at a small angle $\pm \theta$ to the $(\hat{x}-\hat{y})$ direction, the sign being determined by the initial state of the NV center (see Fig.~\ref{fig:polarization}).

In scheme $M2$, we erase quantum information carried by the state-flipping photon by inserting a polarizer along the $(\hat{x}-\hat{y})$ direction in the collection path. In scheme $M3$ we use $(\hat{x}+\hat{y})$ polarizer instead. 

We now analyze scheme $M2$ and come back to scheme $M3$ below. The polarizer only allows photons in the state $\ket{p} = \ket{-} = \frac{-1}{\sqrt{2}}(\ket{\hat{x}}_o-\ket{\hat{y}}_o)$ to reach the detector. Using the relation of the transition amplitudes in Eq.~\eqref{eq:trans_amp_mgcpt}, the transformation of a single NV center state after the detector captures a heralding scattered photon is described by:
\begin{equation}
T_{s} = \frac{\bar{A}}{\sqrt{A_{p}^2+\bar{A}^2}}
\left( \begin{array}{cc}
A_p/\bar{A} & -1\\
1 & - A_p/\bar{A}
\end{array}\right)
\end{equation}
in the basis $\ket{g_2}$ and $\ket{g_3}$, where $\bar{A}$ is the average state-flipping transition amplitude defined as $\bar{A} = (A_1+A_2)/2$. 

Again, assuming the right detector captures a photon, the 2-qubit gate can be described by the matrix,
\begin{equation}
G^{(2)}_{r} =\frac{\bar{A}}{N} \left( \begin{array}{cccc}
\frac{-(1+i)A_{p}}{\bar{A}} & 1 & i & 0 \\
-1 & \frac{(1-i)A_p}{\bar{A}} & 0 & i \\
-i & 0 & \frac{(i-1)A_p}{\bar{A}} & 1\\
0 & -i & -1 & \frac{(1+i)A_p}{\bar{A}} 
\end{array}\right)
\end{equation}
in the basis of $\ket{g_2;g_2}$, $\ket{g_2;g_3}$ and $\ket{g_3;g_2}$ and $\ket{g_3;g_3}$, where the normalization constant is defined as $N^2 = 2 (A_{p}^{2} +\bar{A}^2)$. Note that this gate is still not unitary. The non-unitarity is due to the existence of the residual state-preserving photons that cannot be filtered out from the scattered light. However, since we are working at the ``magic'' frequency of the driving light where the state-preserving transitions are highly suppressed, the gate unitarity is only slightly broken. By the same argument as in Section~\ref{subsec:magic}, with state-preserving transition amplitude $A_p \sim 0.0278$, the entanglement fidelity of this gate is 
\begin{align}
F_{e} = \frac{\bar{A}^2}{\bar{A}^2 + A_{p}^2 }\sim 0.981.
\end{align}

Since the polarization of the state-flipping photon is not aligned to the $(\hat{x}-\hat{y})$ direction exactly, the existence of the polarizer causes the desired photons to have a loss probability, which decreases the gate success probability. In an ideal experimental setup, the gate operation fails if the first state-flipping photon fails to pass the polarizer. Therefore, we calculate the probability that a photon emitted from the NV centers successfully passes the polarizer to estimate the gate success probability. This probability is given by:
\begin{equation}
P_{-} = \bra{-} \textrm{Tr}_{\textrm{NV}} (\rho) \ket{-} = \frac{\bar{A}^{2}+A_{p}^{2}}{\left( A_{1}^{2} + A_{2}^{2}\right)/2 +  A_{p}^{2}}
\label{eq:suc_prob_minus}
\end{equation}
where $\rho$ is the density operator for the NV centers and the scattered photon at the time when the scattering process has occurred but the photon has not gone through the polarizer, $\ket{-} = \frac{1}{\sqrt{2}} \left( -\ket{\hat{x}} + \ket{\hat{y}}\right)$ is the photon state that are allowed to pass the polarizer, $\textrm{Tr}_{\textrm{NV}}$ is the partial trace over all degrees of freedom of NV centers. In this case, the success probability of our gate is $98.1 \%$. 

Scheme $M3$ is similar to scheme $M2$, except that we orient the polarizer along $(\hat{x} + \hat{y})$ direction to only allow photons in state $\ket{p} = \ket{+} = \frac{1}{\sqrt{2}} \left(\hat{x} + \hat{y}\right)$ to pass the polarizer. In this case, the gate is perfectly unitary (when operated at the ``magic'' frequency). Following arguments similar to the $M2$ scheme above, we find that  the 2-qubit gate, conditioned on a click in the right detector, is described by the matrix:
\begin{equation}
G^{(3)}_{r} = \frac{1}{\sqrt{2}}\left( \begin{array}{cccc}
0 & 1 & i & 0 \\ 
1 & 0 & 0 & i \\
i & 0 & 0 & 1\\
0 &  i & 1 & 0
\end{array}\right).
\end{equation}
Note that this gate operation exactly matches Eq.~\eqref{eq:right_detector_gate}. 

However, since the scattered photons from state-flipping transitions are nearly polarized along $(\hat{x} - \hat{y})$ direction, the component along the direction $(\hat{x} + \hat{y})$ is small, which causes a low gate success probability as most state-flipping photons are stopped by the polarizer. Similar to the previous case, the gate success probability is calculated as:
\begin{equation}
P_{+} = \bra{+} \textrm{Tr}_{NV}(\rho) \ket{+} = \frac{\left(A_{1}-A_{2}\right)^{2}/4}{\left( A_{1}^{2} + A_{2}^{2}\right)/2 +  A_{p}^{2}} \sim 1.9 \, \%.
\label{eq:suc_prob_plus}
\end{equation}

\subsection{B1 2-qubit gate scheme: ``Balance'' frequency drive light} \label{subsec:balance_point}
Because of the orthogonality of the dipole moment matrix discussed at the beginning of Section~\ref{subsec:magic}, the scattered photons from state-preserving and state-flipping transitions can be fully distinguished by polarization if the driving light is along $\hat{x}$ or $\hat{y}$ direction. Therefore, besides the ``magic'' frequency of the driving light, we can find a frequency point for the driving light to give us balanced state-flipping transitions and use a polarizer to discard the state-preserving photons. This ``balanced'' point is shown in Fig.~\ref{fig:transition} by the blue arrow. If the driving light is polarized along $\hat{x}$ direction, at the ``balance'' frequency, the state-flipping transition amplitudes satisfy $A_{f,2}^{x} = A_{f,3}^{x}$. Combining this fact with a polarizer along $\hat{y}$ direction in the collection path, if the right detector captures the scattered photon, the 2-qubit unitary gate is described by the matrix
\begin{equation}
G^{(4)}_{r} = \frac{1}{\sqrt{2}}\left( \begin{array}{cccc}
0 & 1 & i & 0 \\ 
1 & 0 & 0 & i \\
i & 0 & 0 & 1\\
0 &  i & 1 & 0
\end{array}\right)
\end{equation}
in the same basis as Eq.~\eqref{eq:gate_1}. 

Unlike in scheme $M3$ that was described in the previous subsection, in scheme $B1$ the state-preserving transition rate is comparable to the state-flipping transition rate. We now point out that the existence of state-preserving transitions, though the scattered photons from these transitions are completely filtered out, decoheres the initial states of the NV centers.

To understand the decoherence mechanism associated with the state-preserving transitions, we construct the master equation to describe the time evolution of the NV center. We assume the NV centers are driven by a $\hat{x}$ polarized light and the polarizer in the collection path is along $\hat{y}$ direction. For simplicity, we assume the emitted photons only couple to the right propagating modes of the waveguide and are detected by the right detector. Since the state-preserving photons are polarized along $\hat{y}$, while the state-flipping photons are polarized along $\hat{x}$, they couple to two different waveguide modes (see~\ref{subsec:waveguide} for details). We further assume the driving light is weak and far-detuned from the excited states, so we can construct an effective Hamiltonian to describe the scattering process where only ground states $\ket{g_2}$ and $\ket{g_3}$ of NV centers appear (see~\ref{subsec:transition_rates} for details). Therefore, we can treat each NV center as a two-level system. We further treat the two waveguide modes as two thermal baths at temperature zero and trace out the photon degrees of freedom, so that the master equation for the NV centers is:
\begin{equation}
\begin{aligned}
\partial_t \rho & = B \left(2\hat{L}\rho \hat{L}^{\dagger} - \hat{L}^{\dagger} \hat{L} \rho - \rho \hat{L}^{\dagger} \hat{L}\right) \\
& + B \left(2 \hat{G} \rho \hat{G}^{\dagger} - \hat{G}^{\dagger} \rho \hat{G} -\hat{G} \rho \hat{G}^{\dagger} \right), \\
\hat{L} & = A_{f,2}^{(x)} \left( i \sigma_{23}^{(1)} + \sigma_{23}^{(2)}\right) + A_{f,3}^{(x)} \left( i \sigma_{32}^{(1)} + \sigma_{32}^{(2)}\right), \\
\hat{G} & = A_{p,2}^{(x)} \left( i \sigma_{22}^{(1)} + \sigma_{22}^{(2)}\right) + A_{p,3}^{(x)} \left( i \sigma_{33}^{(1)} + \sigma_{33}^{(2)}\right),
\end{aligned}
\label{eq:full_master_v1}
\end{equation}
where  $\hat{L}$ and $\hat{G}$ are two jump operators describing the state-flipping transitions and state-preserving transitions respectively, the operator $\sigma_{jk}^{(i)}$ is the operator acting on $i$-th NV center and flips NV state from $\ket{g_j}$ to state $\ket{g_k}$, i.e. $\sigma_{jk}^{(i)} = \ket{g_k}\bra{g_j}$ for $i$-th NV center, and $B = \frac{2 \pi n_{\textrm{eff}}}{c \hbar^2}$ is a constant, where $n_{\textrm{eff}}$ is the mode effective refractive index (see Eq.~\eqref{eq:rate_v1} in~\ref{subsec:transition_rates}). We find that the second term in the master equation involving $\hat{G}$ causes the off-diagonal elements of the two-NV density matrix to decay if the state-preserving transitions are not balanced. This means that if our initial state is prepared in an entangled state of two NV centers, the entanglement between the two NV centers is destroyed by these undetected state-preserving transitions, which will also limit our gate operation time at this frequency point. 

We can also calculate the gate success probability using a similar method to the one illustrated by Eq.~\eqref{eq:suc_prob_minus} and Eq.~\eqref{eq:suc_prob_plus}, which we find to be $37.4 \%$. Note that the success probability is a ``first-photon'' success probability, which means we know in advance the scatter has happened and a single scattered photon has already been emitted into the waveguide mode. In the more realistic case, we can only monitor the detector and we have no information whether the state-preserving transitions happens or not. Gate fidelity and success probability for this case will be discussed in Section.~\ref{sec:fidelity} using quantum trajectory method.

\section{2-qubit gate fidelity and success probability} \label{sec:fidelity}

\begin{figure*}[htbp!]
\includegraphics[width=\textwidth]{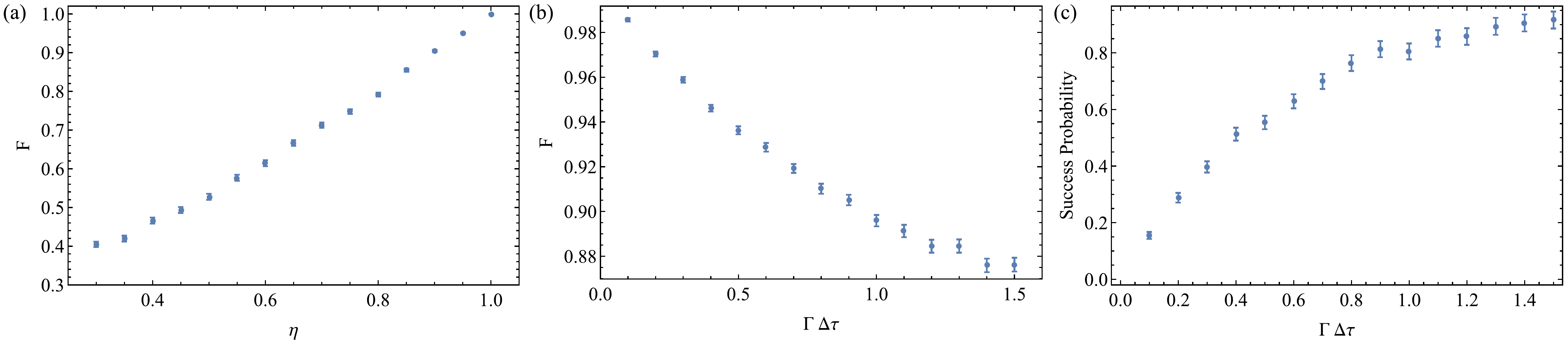}
\centering
\caption{(a) Unitary gate fidelity, $F$, drops as overall scattered photon collection efficiency, $\eta$, decreases with the first gate operation strategy (see main text). The missing scattered photon cases degrade the gate fidelity when photon detection efficiency is imperfect. Using the second gate operation strategy (see main text), the the gate fidelity [in (b)] and success probability [in (c)] is investigated numerically as a function of maximum collection time window ($\Delta \tau$). The overall photon detection efficiency is $85\%$.}
\label{fig:fidqt}
\end{figure*}

In this section, we analyze the fidelity and success probability of our proposed 2-qubit gate for NV centers with possible experimental imperfections. First of all, we notice that NV centers have a phonon side bind which causes Raman scatterings. However, these scattered photons do not have same frequencies as the driving light so that we can filter out and also monitor these photons. The existence of the phonon side band decreases the gate success probability, but does not decrease the gate fidelity. In the following discussion, we ignore the phonon side band and mainly focus on (1) the imperfect scattered photon collection and detection efficiency of the experimental setup, (2) the unbalanced state-flipping transition rates, and (3) possible population loss from the $\ket{g_2}$ and $\ket{g_3}$ manifold. We use quantum trajectory simulations with continuous measurement of the scattered photons to estimate the output state fidelity and success probability with different gate operation schemes and photon collection strategies. In the simulations we use the transition amplitudes calculated at the corresponding driving light frequency and take different types of imperfections together into consideration.

\subsection{Imperfect scattered photon collection and detection efficiency} \label{subsec:imperfect_collection}
Unlike the quantum entanglement proposals in Ref.~\cite{Cabrillo1999, Bose1999, Duan2001, Feng2003, Barrett2005}, when applying a measurement-based unitary gate to two NV centers, in general, we do not know in advance which states these NV centers are. Therefore the NV centers cannot be reset back to initial input state to re-apply the gate operation. It is critical to detect the first state-flipping photon from the two NV centers to perform the unitary gate operation successfully. One possible error source in real experiment for our proposed 2-qubit gate is the imperfect photon collection and detection efficiency of the experimental setup, which we now discuss.

If the detection efficiency of the setup is imperfect, any loss of the heralding photons indicates that undetected state-flipping transitions occurred on either of the two qubits. After missing one or several scattered photons, a photon detection projects the NV centers into an undesired 2-qubit state, which degrades the gate fidelity. To estimate the quality of the gate operation with imperfect photon detection efficiency, we perform quantum trajectory calculation with continuous measurement of the scattered photons to numerically investigate the gate fidelity and success probability. 

In our model, because we only consider the scattering between the states $\ket{g_2}$ and $\ket{g_3}$, we treat NV centers as 2-level systems by using the effective Hamiltonian for the scattering process (see~\ref{subsec:transition_rates} for details). For simplicity, we ignore other imperfections, i.e. our 2-qubit gate is working at a fictitious driving frequency at which two state-preserving transitions are perfectly suppressed and the two state-flipping transitions are balanced. Therefore, the transition amplitudes in Eq.~\eqref{eq:full_master_v1} satisfy $A^{x}_{p,2} = A^{x}_{p,3} = 0$ and $A^{x}_{f,2} = A^{x}_{f,3} \equiv A $ and thus the master equation can be written as,
\begin{equation}
\begin{aligned}
& \frac{\partial \rho}{\partial t} =  -\frac{\Gamma}{2} \left(\hat{L}^{\dagger} \hat{L} \rho + \rho \hat{L}^{\dagger} \hat{L} -2 \hat{L}\rho \hat{L}^{\dagger} \right) \\
& \hat{L} = i \sigma_{23}^{(1)} + \sigma_{23}^{(2)} + i \sigma_{32}^{(1)} + \sigma_{32}^{(2)}
\end{aligned}
\label{eq:masterEq}
\end{equation}
where $\Gamma = B \vert A \vert^2$ is the state-flipping transition rates, $\sigma^{(i)}_{jk} = \ket{g_k}\bra{g_j}$ is the operator for $i$-th NV transiting from state $\ket{g_j}$ to state $\ket{g_k}$ with $j,k = 2,3$. Because in the present consideration, the two state-flipping transitions are balanced, the output state fidelity for all possible input states should be the same and hence the output state fidelity for a certain input state is the gate fidelity. We choose state $\ket{\psi_i} = \ket{g_2} \otimes \ket{g_2}$ as the input state. We labels the 2-NV state $\ket{g_i} \otimes \ket{g_j}$ as $\ket{g_i;g_j}$.

To calculate the output state fidelity of input state $\ket{\psi_i} = \ket{g_2;g_2}$, at the beginning of each trajectory, we initialize both NV centers in $\ket{g_2}$ state and stochastically evolve the two NV centers according to the master equation in Eq.~\eqref{eq:masterEq} conditioned on the measurement result from the detector. When a photon is emitted from NV center, it has probability $\eta$ to be detected by the detector, otherwise the photon is lost into the bath. The photon detection is a projection measurement, with the jump operator $\hat{L}$ in Eq.~\eqref{eq:masterEq} as the measurement projector. When a scattered photon is detected by the detector, the density matrix collapses to $\rho' \propto \hat{L}\rho\hat{L}^{\dagger}$ up to a normalization constant. It is obvious that if the detection efficiency $\eta=1$, the gate operation is a 2-qubit unitary gate described by $G_r$ in Eq.~\eqref{eq:right_detector_gate}.

The first strategy to perform the 2-qubit unitary gate is to run the trajectory until we receive a photon by the detector. In real experiment, it is equivalent to running the experiments until a photon is detected without limiting the collection time window. When a photon is detected, we stop the time evolution of the trajectory and calculate the output state fidelity using the target state $\ket{\psi_{\text{t}}} = G_r \ket{\psi_i} = \frac{1}{\sqrt{2}} \left( \ket{g_2;g_3} + i \ket{g_3;g_2}\right)$. Since we do not limit the total time to end the protocol, we always have a positive detection result and thus the gate is always considered as success. However, the gate fidelity suffers from the missing photon cases. We ran $1000$ independent trajectories in total to build up statistics for the gate fidelity. The gate fidelity as a function of overall photon detection efficiency ($\eta$) is shown in Fig.~\ref{fig:fidqt}(a). The numerical simulation matches our expectation that as the collection efficiency drops, it becomes more and more likely that the first scattered photon is missed, and hence the overall output state fidelity drops. When the collection efficiency  $\eta=1$, the fidelity is $1$. The fidelity drops to $0.5$ when the overall photon detection efficiency drops to $\eta \sim 0.45$. Based on the proposed geometry of the diamond waveguide, we calculate the overall collection efficiency of the diamond waveguide to be $85 \%$ (see Appendix for details). At the $85 \%$ photon collection efficiency, the gate fidelity is $0.8547 \pm 0.0040$. 

The second strategy aims to improve gate fidelity with an imperfect photon detection efficiency, by limiting the maximum photon collection time window. This will help to rule out missing photon cases and improve the fidelity of the 2-qubit gate operation. However, as we decrease the collection window, it is possible not to detect any photons within the time bin, and hence the gate success probability is expected to drop as we shrink the collection window. We numerically investigate the output state fidelity and success probability as we change the duration of collection window. We use the same quantum trajectory method with a collection efficiency $\eta$ to stochastically time evolve the master equation in Eq.~\eqref{eq:masterEq}. We still use the state $\ket{\psi_i} = \ket{g_2;g_2}$ as the input state and $\ket{\psi_t} = \frac{1}{\sqrt{2}} (\ket{g_2;g_3} + i \ket{g_3;g_2})$ as the target state. If we get a positive detection result within the collection window, we stop the trajectory and measure the output state fidelity. Otherwise, if no scattered photon is detected till the end of the collection window, we reckon the gate fails and stop the trajectory. The numerically calculated average gate fidelity and gate success probability with $\eta=0.85$ as we change the collection window is plotted in Fig.~\ref{fig:fidqt}(b) and Fig.~\ref{fig:fidqt}(c) respectively. The average gate fidelity improves as we shrink the collection window, but the success probability drops, as we expected. For example, if we choose the collection window $\Gamma \Delta \tau = 0.1$, the fidelity can be improved to $0.9857 \pm 0.0007$, however, the success probability of the gate decreases to $0.155$. To conclude, this gate operation strategy trades the successful probability for high gate fidelity.

We want to point out that Ref.~\cite{Benjamin2009} shows that constructing a graph or cluster state requires a minimum success probability of $1/3$. In our numerical simulations this threshold can be met by setting the collection window to be $\Gamma\Delta \tau=0.3$, which results in the gate success probability of $0.397$ and an average output state fidelity of $0.9588 \pm 0.0013$.

\subsection{Unbalanced state-flipping transitions} \label{subsec:unbalance_transition}
In the above calculation, we assumed that the two state-flipping transition rates are balanced. However, this assumption does not have to hold. For example, in scheme $M1$, which we discuss in Section~\ref{subsec:magic}, the transition rates for the two state-flipping transitions are different. Furthermore, the state-flipping transition rates of two NV centers may also be different (e.g. due to different coupling strength to the waveguide modes). In Section~\ref{subsec:magic}, we considered the gate fidelity when the state-flipping transitions rates are not equal, but two NV centers are identical. In this subsection we consider a more general case when the two state-flipping transitions of two NV centers emit indistinguishable scattered photons, but the rates can be different. We analyze the gate operation and the gate fidelity.

When the state-flipping transition rates are different from one NV center to the other one, we use $A^{(i)}_{1}$ and $A^{(i)}_{2}$ to note the transition amplitude for state-flipping transitions from $\ket{g_2}$ to $\ket{g_3}$ and $\ket{g_3}$ to $\ket{g_2}$ of $i$-th NV center. Here we assume there is no state-preserving transitions and detection efficiency is $1$ to only focus on the imperfection caused by the unbalanced state-preserving transitions. We also assume the state-flipping transition amplitudes are all positive. 

Similar to the previous subsection, we assume the scattered photons only couples to the right-propagating modes, and thus the master equation of the two NV centers in this case is similar to the master equation shown in Eq.~\eqref{eq:full_master_v1} as,
\begin{equation}
\begin{aligned}
& \frac{\partial \rho}{\partial t} =  -\frac{B}{2} \left(\hat{L}^{\dagger} \hat{L} \rho + \rho \hat{L}^{\dagger} \hat{L} -2 \hat{L}\rho \hat{L}^{\dagger} \right) \\
& \hat{L} = i A^{(1)}_{1} \sigma_{23}^{(1)} + A^{(2)}_{1} \sigma_{23}^{(2)} + i A^{(1)}_{2} \sigma_{32}^{(1)} + A^{(2)}_{2} \sigma_{32}^{(2)}
\end{aligned}
\end{equation}
When a photon is captured by the detector, it corresponded to a projection measurement onto the NV centers which is described by the jump operator $\hat{L}$. Therefore the gate operation can be described by the matrix,
\begin{equation}
\hat{L} = \left( \begin{array}{cccc}
0 & A^{(2)}_2 & i A^{(1)}_2 &  0\\
A^{(2)}_1 & 0 & 0 & i A^{(1)}_2 \\
i A^{(1)}_{1} & 0 & 0 & A^{(2)}_2\\
0 & i A^{(1)}_{1} & A^{(2)}_1 & 0\\
\end{array}
\right)
\end{equation}
in the same basis as Eq.~\eqref{eq:gate_1}. We can use the same method as discussed in Section~\ref{subsec:magic} to estimate the gate fidelity. We can define $\bar{A}$ as the average of these four state-flipping transition amplitudes as $\bar{A} = \sum_{i,j} A^{(i)}_{j}/4$ and the derivations of each specific transition amplitude from this average amplitude by $\delta_{i,j} = A^{(i)}_j -\bar{A}$. When the four transition amplitudes are not severely unbalanced, i.e. $\left\vert \delta_{i,j}/\bar{A}\right\vert \ll 1$, we can expand the output state fidelity in series of $\delta_{i,j}/\bar{A}$. In general, the gate fidelity will drop linearly as $\delta^{2}_{i,j}/\bar{A}^2$ increases. As we see from Section~\ref{subsec:magic}, when $A^{(1)}_i = A^{(2)}_i$, the deviation of the transition amplitudes $\delta_{i,1} = - \delta_{i,2} \equiv \delta$. The gate fidelity can then be expanded as,
\begin{equation}
F = \frac{\bar{A}^2}{\bar{A}^2 + \delta^2} \sim 1 - \frac{\delta^2}{\bar{A}^2}
\label{eq:expand_fid}
\end{equation}

Let's also discuss the case when two state-flipping transition amplitudes for a single NV center are balanced, however, the same transitions for different NV centers have a constant transition amplitude offset. In this case, we assume $A^{(1)}_{j} = \bar{A} - \delta$, and $A^{(2)}_{j} = \bar{A}+\delta$. The gate fidelity is also given by Eq.~\eqref{eq:expand_fid}. 

\subsection{Overall output state fidelity}
In this subsection, we evaluate the gate quality by numerically simulating the output state fidelity and success probability with the four possible gate operation schemes discussed in Section~\ref{sec:transitions} combined with the two proposed collection strategies discussed in Section~\ref{subsec:imperfect_collection}. The four gate operation schemes are summarized in Table.~\ref{tab:schemes}. The two collection strategies are collecting the photon (1) without and (2) with a maximum collection window $\Delta \tau$.
 
With all four gate operation schemes, we explore the output state fidelity when state $\ket{\psi_1} = \ket{g_2;g_2}$, $\ket{\psi_2}=\ket{g_2;g_3}$ and $\ket{\psi_3} = \frac{1}{\sqrt{2}} \left( \ket{g_2;g_2} + i \ket{g_3;g_3}\right)$ as the gate input states using quantum trajectory simulation with continuous measurement on the scattered photons. We set the overall collection efficiency of the photons through the polarizer to $85 \%$. The gate average fidelity and gate success probability without and with a maximum collection time window $\Delta \tau = 0.1 / \bar{\Gamma}_f$ is shown in Table~\ref{tab:fid_list}. Here, $\bar{\Gamma}_f$ is the average state-flipping transition rates, $\bar{\Gamma}_f = \left( A_{1}^{2} + A_{2}^{2}\right)/2$, where $A_{1}$ and $A_{2}$ is the absolute value of the state-flipping transition amplitudes at the working frequency [see Eq.~\eqref{eq:trans_amp}]. We also listed the output state fidelity with corresponding gate operation schemes with perfect photon detection efficiency and infinite pump power for reference, which set a theoretical upper bound for the output state fidelity in the corresponding cases. 

\begin{table}[tbp]
\caption{\label{tab:fid_list} 
Output state fidelity and gate success probability for input states $\ket{\Psi_1} = \ket{g_2;g_2}$, $\ket{\Psi_2} = \ket{g_2;g_3}$, $\ket{\Psi_3} = \frac{1}{\sqrt{2}} (\ket{g_2;g_2}+i\ket{g_3;g_3})$ with the four gate operation schemes, $M1$, $M2$, $M3$ and $B1$ (see Table.~\ref{tab:schemes}), when the photon collection efficiency is perfect (labeled Perfect Collection), imperfect with an infinite photon collection time window (labeled $\eta = 0.85, \bar{\Gamma}_f \Delta \tau = \infty$), and imperfect with a finite photon collection time window (labeled $\eta = 0.85, \bar{\Gamma}_f \Delta \tau = 0.1$). Note for the case of perfect collection, and the case of imperfect collection with infinite photon collection time window $P=1$.}
\begin{ruledtabular}
\begin{tabular}{l|c|c|cc}
Input &  Perfect & $\eta=0.85$ & \multicolumn{2}{c}{$\eta = 0.85$} \\
State & Collection &  $\bar{\Gamma}_f \Delta \tau = \infty$ & \multicolumn{2}{c}{$\bar{\Gamma}_f \Delta \tau = 0.1$} \\
& $F$ & $F$ &  $F$ & $P$ \\
\hline
$M1$ & & & \\
$\ket{\Psi_{1}}$ & $1.0$ & $0.848 \pm 0.004 $ &  $0.9896 \pm 0.0006 $ & $0.106$\\
$\ket{\Psi_{2}}$ & $0.981$ & $0.837 \pm 0.005$  & $ 0.9704 \pm 0.0005$ & $0.164$ \\
$\ket{\Psi_{3}}$ & $0.981$ & $0.824\pm 0.005$  & $ 0.9665 \pm 0.0006$ & $0.156$ \\
\colrule
$M2$ & & &\\
$\ket{\Psi_{1}}$ & $0.981$ & $0.819 \pm 0.005 $ &  $0.9683 \pm 0.0006$ & $0.172$ \\
$\ket{\Psi_{2}}$ & $0.981$ & $0.824 \pm 0.005 $ &  $0.9678 \pm 0.0006 $ & $0.166$\\
$\ket{\Psi_{3}}$ & $0.981$ & $0.823 \pm 0.005 $ &  $0.9683 \pm 0.0006 $ & $0.169$\\
\colrule
$M3$ & & &\\
$\ket{\Psi_{1}}$ & $1.0$ & $0.255 \pm 0.002 $ &  $0.916 \pm 0.004 $ & $0.0037$\\
$\ket{\Psi_{2}}$ & $1.0$ & $0.256 \pm 0.002 $ & $0.902 \pm 0.004$ & $0.0035$ \\
$\ket{\Psi_{3}}$ & $1.0$ & $0.255 \pm 0.002$  & $ 0.911 \pm 0.004$ & $0.0033$ \\ 
\colrule
$B1$ & & &\\
$\ket{\Psi_{1}}$ & $ 1.0 $ & $0.859 \pm 0.004 $ &  $0.9870 \pm 0.0006 $ & $0.172$\\
$\ket{\Psi_{2}}$ & $ 1.0 $ & $0.857 \pm 0.004 $ &  $0.9842 \pm 0.0007 $ & $0.153$\\
$\ket{\Psi_{3}}$ & $ 1.0 $ & $0.571 \pm 0.006 $ &  $0.906 \pm 0.004 $ & $0.150$\\
\end{tabular}
\end{ruledtabular}

\end{table}

To estimate the gate fidelity of the different schemes we use the worst output state fidelity in Table~\ref{tab:fid_list}. $M3$ and $B1$ are two schemes that are perfectly unitary in ideal conditions. When we don't setup a finite collection window, since the gate operation scheme $M3$ suffers low success probability, even with perfect collection efficiency, the output state fidelity drops significantly from unity. This is because most of the detected photons are from the long-time scatter events, i.e. the NV center system tends to relax to its steady state before the heralding photon is detected. Therefore, it is equivalent to applying the gate to the steady state of the master equation, which gives an output state fidelity 
$\approx 0.25$. If we don't limit the collection window, the gate operation scheme $B1$ has significantly different output state fidelity when the input state is $\ket{\psi_1}$ (or $\ket{\psi_2}$) and $\ket{\psi_3}$. This is because the undetected state-preserving transitions decohere the input state, even though they do not flip the NV spin states and their photons are perfectly separated from the state-flipping photons. The input state $\ket{\psi_3}$ decoheres to an equal mixture of states $\ket{g_2;g_2}$ and $\ket{g_3;g_3}$, which makes the output state-fidelity drop to $\approx 0.5$. The finite collection time window helps to discard the long-time detection events, which improves the output-state fidelity significantly, especially for the gate operation scheme $M3$.

Gate operation schemes $M1$ and $M2$ are not perfectly unitary even in the ideal case. However, since the polarizer setup has little probability to block the state-flipping photons and the state-preserving transitions are highly suppressed due to the ``magic'' frequency of the driving light, these two schemes behave much better when the collection time is not limited. When we have a finite collection window, the output state fidelity also improves. Compared to the gate operation schemes $M3$ and $B1$, the schemes $M1$ and $M2$ have better output state fidelity.

\subsection{Population loss due to the transition out of the $\ket{g_2}$, $\ket{g_3}$ manifold}
\begin{figure}[b]
\includegraphics[width=3.2 in]{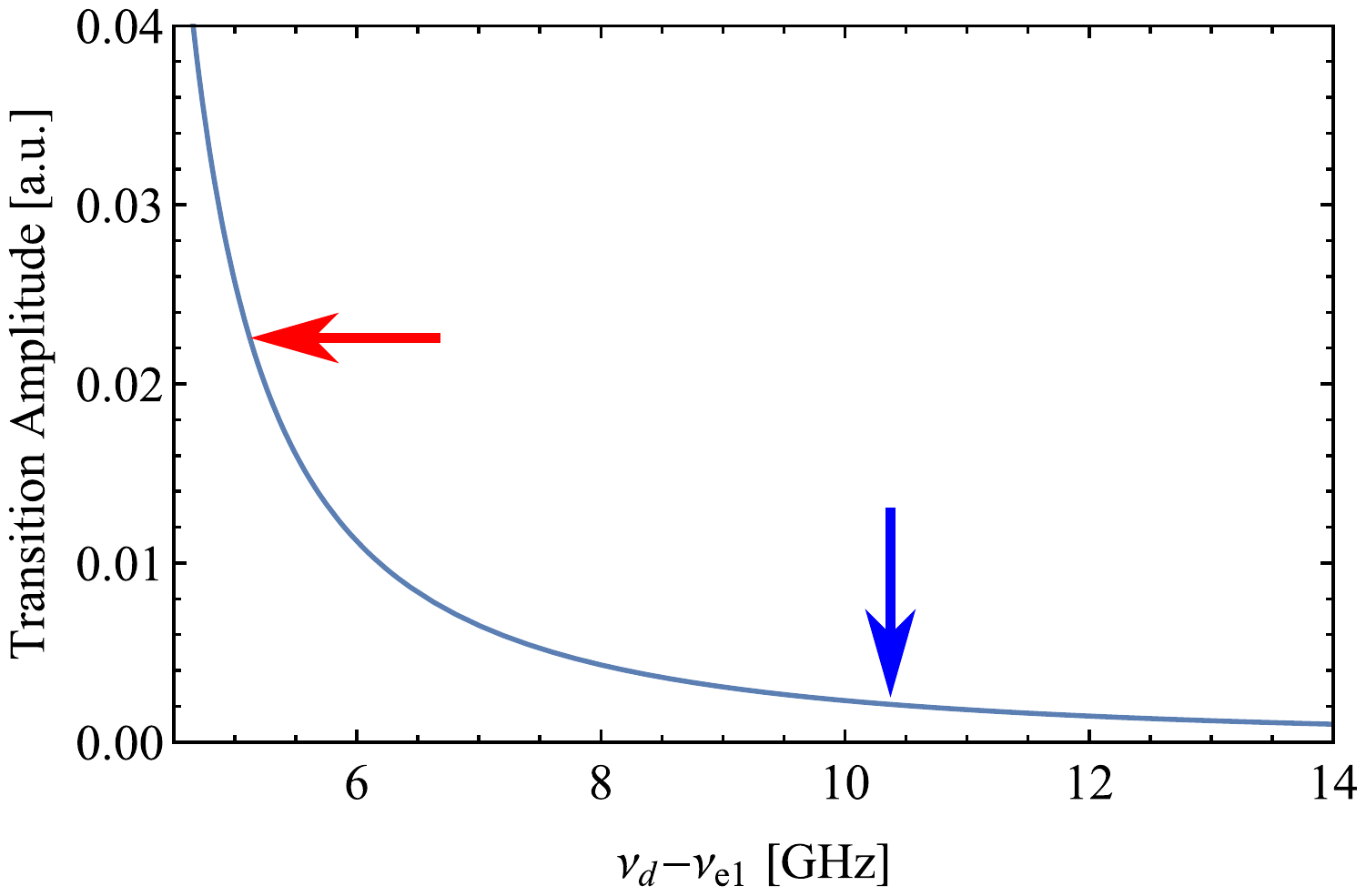}
\centering
\caption{We study the population leakage from state $\ket{g_2}$ and $\ket{g_3}$ manifold into state $\ket{g_1}$ state caused by spin-Raman transitions. We plot the transition rate under the same pumping laser as in Fig.~\ref{fig:transition}. The ``magic'' frequency is pointed out by the red arrow while the ``balance'' frequency is labeled by the blue arrow. The population leakage rate by spin-Raman transition is much slower than the state-flipping transitions shown in Fig.~\ref{fig:transition} and hence we do not expect to see large population within the detection window.}
\label{fig:leakage}
\end{figure}

Any process that transfers population out of $\ket{g_2}$ and $\ket{g_3}$ manifold, i.e. to the other states like $\ket{g_1}$, results in no further photon detections after this ``leakage'' transition happens. This will degrade the success probability of the gate. There are two possible leakage paths, (1) by the Raman scattering process to state $\ket{g_1}$, (2) by exciting to the NV electronic excited states then by the non-radiative relaxation through the meta-stable states of NV centers to $\ket{g_1}$.

To examine the effect of spin Raman transition from logic states $\ket{g_2}$ and $\ket{g_3}$ to state $\ket{g_1}$, we refer to the dipole matrix in Eq.~\eqref{eq:dipole_SS} in~\ref{subsec:dipole}, and calculate the leakage transition amplitudes as,
\begin{equation}
\begin{aligned}
\frac{A^{x}_{l,2}}{A^{(x)}_{0}} & = \frac{1}{\Delta_1} F_{21}F_{11} - \frac{1}{\Delta_3} F_{22}F_{12} \\
\frac{A^{y}_{l,2}}{A^{(y)}_{0}} & = - \frac{1}{\Delta_1} F_{21}F_{11} + \frac{1}{\Delta_3} F_{22}F_{12} \\
\frac{A^{x}_{l,3}}{A^{(y)}_{0}} & = - \frac{1}{\Delta_2} F_{21}F_{11} + \frac{1}{\Delta_4}F_{22}F_{12} \\
\frac{A^{y}_{l,3}}{A^{(x)}_{0}} & = - \frac{1}{\Delta_1} F_{21}F_{11} + \frac{1}{\Delta_3}F_{22}F_{12}
\end{aligned}
\label{eq:leak_amp}
\end{equation}
Where $F_{11} = 0.0513$ and $F_{12} = 0.9987$ are two dimensionless parameters from the dipole moments between eigenstates of spin-orbit and spin-spin Hamiltonian of single NV centers (see Eq.~\eqref{eq:dipole_SS} in~\ref{subsec:dipole}), $\Delta_i$ are the energy mismatch for excited level $\ket{e_i}$. If we consider the fact that the excited states $\ket{e_1}$ and $\ket{e_2}$, $\ket{e_3}$ and $\ket{e_4}$ are energetically degenerate, i.e. $\Delta_1 = \Delta_2$, $\Delta_3 = \Delta_4$,  these four transition amplitudes satisfies $ -\frac{A^{x}_{l,2}}{A^{(x)}_{0}} = \frac{A^{y}_{l,2}}{A^{(y)}_{0}} = \frac{A^{x}_{l,3}}{A^{(y)}_{0}} = \frac{A^{y}_{l,3}}{A^{(x)}_{0}}$. We plot the magnitude of R.H.S of Eq.~\eqref{eq:leak_amp} in Fig.~\ref{fig:leakage}, and label the ``magic'' point and ``balance'' point by red and blue arrows respectively. At the ``balance'' point, the leak transition amplitudes are two orders of magnitudes smaller than the state-flipping transition amplitudes and hence have little impact on the gate operation scheme $B1$. The population of the NV centers in ground states $\ket{g_2}$ and $\ket{g_3}$ decays slowly to $\ket{g_1}$ due to the existence of the leakage transitions, which sets a maximum gate operation window to avoid significant population loss.

At the ``magic'' point, the leak transition amplitudes are comparable to the state-preserving transition amplitudes. Note that this suppression is not due to the interference. Instead, it is mainly suppressed by the small mixing of excited spin $S_z=0$ states with spin $S_z=\pm 1$ states that caused by the spin-spin interaction~\cite{Doherty2011}. Compared to the state-flipping transition amplitudes, the leakage transition amplitudes are approximately ten times smaller than the state-flipping transition amplitudes. The gate operation schemes working at the ``magic'' frequencies are not severely affected.

To quantitatively estimate the effect of the non-radiative relaxation process, we approximate the dynamics of NV centers with the metastable spin-singlet states as a three-level system, ground state $\ket{0}$, excited state $\ket{1}$ and meta-stable state $\ket{2}$. The transition between states $\ket{0}$ and $\ket{1}$ are driven by an off-resonance classical laser field. The non-radiative relaxation process from state $\ket{1}$ to meta-stable state $\ket{2}$ are modeled by the coupling to a thermal optical phonon bath with temperature zero. Therefore the dynamics can be described by the master equation
\begin{equation}
\begin{aligned}
\partial_t \rho & = -i (2\pi) \left[ -\delta \sigma_{00} + \Omega_{R} \left(\sigma_{01}+ \sigma_{10} \right), \, \rho \right] + \mathcal{L} \rho, \\
\mathcal{L}\rho & = -\frac{\Gamma_{NR}}{2} \left( \sigma_{11}\rho + \rho\sigma_{11} - 2 \sigma_{21}\rho\sigma_{12} \right),
\end{aligned}
\label{eq:leakage_master_equation}
\end{equation}
where operators $\sigma_{ij}$ are defined by $\sigma_{ij} = \vert i \rangle \langle j \vert$, $h\delta = \epsilon_{1} - \epsilon_{0} - h \nu_{d}$ is the detuning of the drive field, $\epsilon_{i}$ is the energy of the state $\ket{i}$, $h \Omega_{R} = p_0 E_{d}$ is the Rabi frequency, $p_0$ is the dipole moment for the optical transition between $\ket{0}$ and $\ket{1}$, which is approximated as $p_0 \approx 5.2$~Debye (see~\ref{subsec:transition_rates} and Ref.~\cite{Alkauskas2014}), $E_{d}$ is the driving light electric field, $\Gamma_{\textrm{NR}}$ is the non-radiative relaxation rate from state $\ket{1}$ to $\ket{2}$.

We estimate the non-radiative relaxation rate $\Gamma_{\textrm{NR}}$ by the lifetime of the excited levels of NV centers. In Ref.~\cite{Doherty2013}, a six-level model is introduced to describe the NV center electronic structure. The excited manifold is simplified as two states with quantum number $S_z = 0$ and $S_z = \pm 1$, with measured lifetime $12.0$~ns and $7.8$~ns respectively~\cite{Batalov2008}. We further assume that the excited state $S_z = 0$ has no relaxation path to the meta-stable state and the radiative relaxation from excited states back to ground states of NV centers are the same, and hence the non-radiative relaxation rate from excited state $S_z = \pm 1$ can be estimated using the difference of the lifetimes of these two excited states as $\Gamma_{\textrm{NR}} \approx 44.9$~MHz. 

We approximate the detuning by the smallest detuning of our driving light, to one of the four excited states with $S_z \sim \pm 1$, i.e. $\ket{e_1}$, $\ket{e_2}$, $\ket{e_5}$ and $\ket{e_6}$. If our proposed gate is working at the ``magic'' frequency of the driving light, the detuning $\delta \approx 3.95$~GHz for a $\hat{y}$ polarized driving light and $5.11$~GHz for a $\hat{x}$ polarized driving light. Clearly, $\Gamma_{\textrm{NR}} / \delta \ll 1$, so that we work in the dressed-state basis and then treat the Lindblad term $\mathcal{L}\rho$ in Eq.~\eqref{eq:leakage_master_equation} as a perturbation. 

In our previous treatment of scattering transitions, we implicitly assumed that the Rabi frequency is small compared to detuning, i.e. $\Omega_{R} / \delta \ll 1$. The dressed state basis for the Hamiltonian in Eq.~\eqref{eq:leakage_master_equation} is $\ket{-} \sim \ket{0} - \frac{\Omega_{R}}{\delta} \ket{1}$ and $\ket{+} \sim \ket{1} + \frac{\Omega_{R}}{\delta} \ket{0}$. If all the population is in state $\ket{0}$ at the beginning, we would expect most of the population will be remain in the state $\ket{-}$ after we start driving the Rabi oscillation. Since the non-radiative relaxation removes the population in state $\ket{1}$ only, the decay rate for the population in state $\ket{-}$ is $\Gamma_{-} \sim \Gamma_{\textrm{NR}} \sigma_{11}\ket{-}\bra{-}\sigma_{11} \sim \Gamma_{\textrm{NR}} \frac{\Omega_{R}^2}{\delta^2} \propto E_{d}^2$. As we show in~\ref{subsec:transition_rates}, the state-flipping transition rate at the ``magic'' point is $\Gamma_{t} \sim \Gamma_0 \propto E_{d}^2$, we can calculate the ratio between the lower state-flipping transition rates versus the non-radiative relaxation rate as $\Gamma_{t} / \Gamma_{-} \sim 1.63$ and $0.975$ for $\hat{x}$ and $\hat{y}$ polarized driving light respectively, which are independent of the driving strength $E_{d}$. These two ratios set a hard limit on the collection time window of the scattered photon before the population is lost.

We perform the same calculation at the ``balance'' point,  and determine the hard limit on the collection window. As the ``balance'' point is located between the excited states $\ket{e_5}$ and $\ket{e_6}$, this balance  frequency for gate operation is more vulnerable to population loss. The transition ratio $\Gamma_{t}/\Gamma_{-}$ is calculated as $0.744$ and $0.412$ for $\hat{x}$ and $\hat{y}$ polarized driving light at ``balance'' point. We summarize the parameters we used and the results in Table~\ref{tab:data} for reference.

\begin{table*}[htbp!]
\caption{\label{tab:data} Summary of the parameters we used for estimating the effect of the non-radiative relaxations. We also listed the smallest frequency detuning when the drive light is at the ``magic'' frequency and the ``balanced'' frequency and the corresponding ratio between lower state-flipping transition rate versus the non-radiative relaxation rate, $\Gamma_{t}/ \Gamma_{-}$.}
\begin{ruledtabular}
\begin{tabular}{l l c}
NV-center electronic dipole moment & $p_0$ & $5.2$~Debye \\
Non-radiative relaxation rate for NV excited states & $\Gamma_{\textrm{NR}}$ & $44.9$~MHz \\
$\hat{x}$-polarized drive at ``magic'' frequency & detuning $\delta$ & $5.11$~GHz \\
& transition rates ratio $\Gamma_t/\Gamma_{-}$ & 1.63\\
$\hat{y}$-polarized drive at ``magic'' frequency & detuning $\delta$ & $3.95$~GHz \\
& transition rates ratio $\Gamma_t/\Gamma_{-}$ & 0.975\\
$\hat{x}$-polarized drive at ``balance'' frequency & detuning $\delta$ & $3.45$~GHz \\
& transition rates ratio $\Gamma_t/\Gamma_{-}$ & 0.744\\
$\hat{y}$-polarized drive at ``balance'' frequency & detuning $\delta$ & $2.57$~GHz \\
& transition rates ratio $\Gamma_t/\Gamma_{-}$ & 0.412\\
\end{tabular}
\end{ruledtabular}
\end{table*}

\section{Summary and Outlook} \label{sec:summary}
In this paper, we proposed a 2-qubit unitary quantum gate to achieve quantum logic operations using two NV centers. We theoretically analyzed how a photon is scattered by an NV center, taking care of the interference between different excitation paths. We found that for scattering rates between two electronic spin states ($\ket{S_z=\pm1}$) there are two special frequencies for the driving light: a ``magic'' frequency at which the state conserving scattering rate is suppressed and a ``balanced'' frequency at which the state-flipping transition rates are equal. We analyzed the gate unitarity, fidelity and success probability for each of the schemes with possible experimental imperfections.  When the photon collection efficiency is $\sim 0.85$, the gate fidelity of the most reliable scheme can reach $\sim 0.97$ when we impose a photon collection window $0.1/\bar{\Gamma}_{f}$, where $\bar{\Gamma}_{f}$ is the averaged state-flipping transition rate. While decreasing the photon collection window can improve the gate fidelity, the corresponding decrease in the success probability will have to be mitigated by some other means to ensure we stay above the threshold for cluster or graph-state quantum computing. The proposed scheme could also be extended to other qubits such as Silicon-vacancy in diamond, or to localized vibronic states of the NV or other defect centers where the larger energy splittings can allow for quantum computing even at room temperature.

\section{Acknowledgment}
The authors acknowledge useful discussion with Roger Mong and Sophia Economou. The work was supported by the Charles E. Kaufman foundation Grant Number KA2014-73919 (CL, MVGD, DP), ARO (CL, DP), NSF Grant Number EFRI ACQUIRE 1741656 (MVGD).

\section{Appendix} \label{Apendix}
\renewcommand{\thesection}{}
\renewcommand{\thesubsection}{Appendix \Alph{subsection}}

\subsection{Waveguide modes and the NV center coupling strength} \label{subsec:waveguide}

\begin{figure*}[htbp!]
\includegraphics[width=7 in]{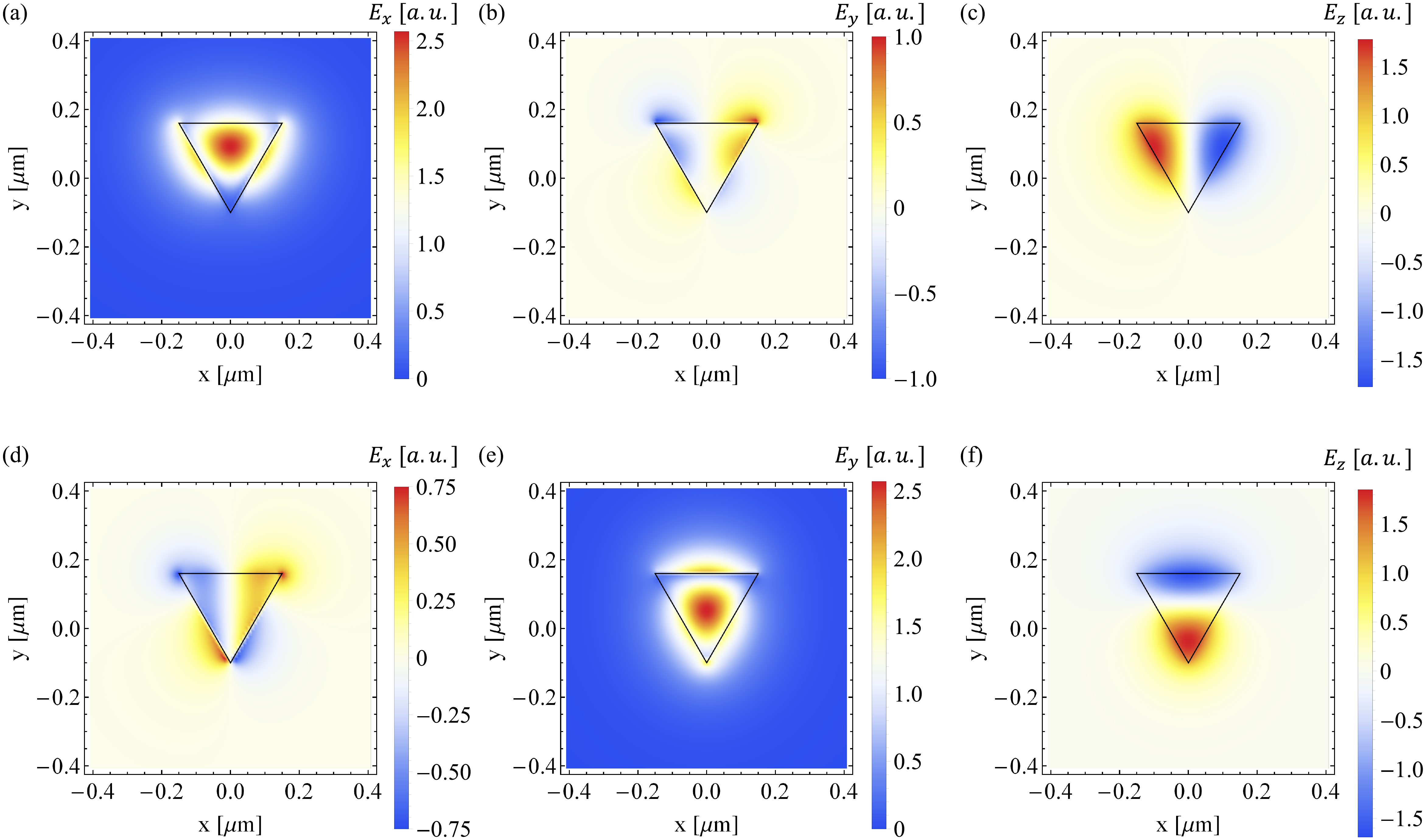}
\caption{The mode profiles of the triangular diamond waveguide. The waveguide has $300$ nm edge. The diamond waveguide supports two degenerate propagating modes. Mode 1 $E_x$, $E_y$ and $E_z$ components are plotted in (a) to (c), while mode 2 components are plotted in (d) to (f).}
\label{fig:waveguide}
\end{figure*}

In this section of the appendix, we analyze the triangular diamond waveguide and its mode profiles. The triangular diamond waveguide we proposed in our paper has $300$ nm edge. The diamond waveguide can be experimentally fabricate using anisotropic plasma etching~\cite{Burek2012}. The mode profiles are calculated by solving eigenproblem of discretized transverse Maxwell equation using Lumerical Mode solution solver. There are only two degenerate guided modes at the ``magic'' frequency. The mode profiles are shown in Fig.~\ref{fig:waveguide}. The modes are normalized according to,
\begin{equation}
\int d x d y \epsilon_{r} (x,y) \vec{E}_{m}^{*} (x,y) \cdot \vec{E}_{n} (x,y) = \delta_{m,n}
\label{eq:mode_normalization}
\end{equation}
where indices $m$ and $n$ are for modes, $\epsilon_r$ is the relative permittivity.

To calculate the light collection efficiency of the diamond waveguide, we treat the NV-center as a dipole moment $\vec{p} = \lvert p \rvert \cdot \hat{p}$ located at position $\vec{r}_0$, where $\hat{p}$ is the unit vector along the dipole moment. We only consider the dipole interaction between NV-centers and the modes inside the waveguide. If we have a well defined mode in the cross-section, whose electric field is $\vec{E}_n(\vec{r})$, the emission rate from the NV-center to this mode $\Gamma_n$ is proportional to $\lvert p \rvert^2 \cdot \lvert \vec{E}_n(\vec{r}_0) \cdot \hat{p} \rvert^2$. For a complete set of orthonormal modes in space with frequency of emission light $\{ \vec{E}_n(\vec{r}) \}$, the total rate can be calculated as $\Gamma_{\text{total}} = \sum_n \lvert p \rvert^2 \cdot \lvert \vec{E}_n(\vec{r}_0) \cdot \hat{p} \rvert^2$. Therefore, the collection efficiency of the waveguide is,
\begin{equation}
\eta(\vec{r}_0)=\frac{ \sum_{n}^{\prime} \lvert \vec{E}_n(\vec{r}_0) \cdot \hat{p} \rvert^2}{\sum_{n}  \lvert \vec{E}_n(\vec{r}_0) \cdot \hat{p} \rvert^2},
\label{eq:collection}
\end{equation}
where $\sum_{n}^{\prime}$ is the summation over the guided modes only, an $\sum_n$ is the summation over all the modes in the complete set of orthonormal modes.

\begin{figure*}[htbp!]
\includegraphics[width=\textwidth]{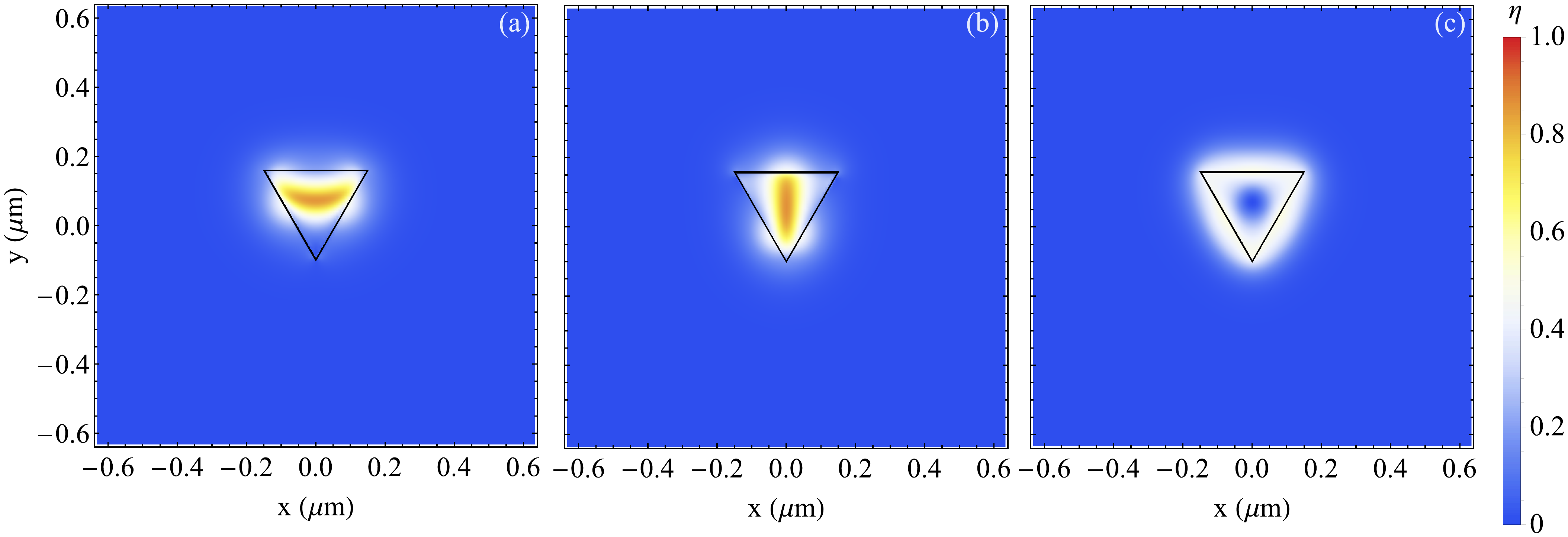}
\caption{The diamond waveguide collection efficiency of the Raman photon emitted from a NV-center located in the cross-section of the waveguide. The NV-center is modeled as a dipole moment. The black triangle labeled in the plot shows the diamond waveguide boundary. The collection efficiency of  photons when the dipole moment is pointing along $x$, $y$ and $z$ direction is plotted in (a), (b) and (c).}
\label{fig:collection}
\end{figure*}

In the numerical approach, we cannot solve an infinite large region. Instead, we solve the modes using a finite size cross-section region. The boundary condition around the region is chosen as perfect matched layer (PML) to simulate the infinite space. We plot the collection efficiency of the diamond waveguide with a dipole moment pointing along $x$, $y$ and $z$ direction at different position in this cross-section in Fig.~\ref{fig:collection}. From the figure, the collection efficiency for a NV-center whose electric dipole moment is along the $x$ or $y$ direction is $\eta \approx 0.86$. However, when the dipole moment is pointing along $z$ direction, the collection efficiency is poor because a dipole moment pointing along $z$ direction mainly radiates in a direction transverse to the direction of the waveguide.

Assuming the NV center is centered in the waveguide, i.e. $x,y \sim 0$, and the NV center is orientated as Fig.~\ref{fig:MBEphonon}(c) shows, the optical dipole moment is along the transverse direction of the waveguide. According to Fig.~\ref{fig:collection}, the NV center optical transitions with $\hat{x}$ dipole moment strongly couples to the mode~1 and almost no coupling to mode~2, while the transitions with $\hat{y}$ dipole moment strongly couples to the mode~2 and almost no coupling to mode~1.

\subsection{Dipole moment of NV-centers without external magnetic field} \label{subsec:dipole}
In this section, we discuss the NV center dipole moment matrix for optical transitions between electronic ground and excited state of NV centers with spin-orbit, spin-spin interactions, and with strain field in diamond crystal. We assume there is no magnetic field applied to the NV center. Here, we follow the notation of Ref.~\citep{Doherty2011}, which gives a detailed review of the electronic properties of negatively charged NV centers. We want to stress that the directions $\hat{x}$, $\hat{y}$ and $\hat{z}$ in this section are the intrinsic directions of an NV center. The direction $\hat{z}$ is defined as the axial direction of NV center, i.e. the direction along the nitrogen atom and the vacancy site, which is the $[111]$ direction of the diamond crystal.

The NV center electronic fine states structure is shown in Fig.~\ref{fig:transition}(a) of our main paper. Here we assume the dipole moment operator $\hat{\vec{p}}$ between the molecule orbits of NV-centers are,
\begin{equation}
\bra{e_x}\hat{\vec{p}}\ket{a_1} = p_0 \cdot \hat{x} \quad , \quad  \bra{e_y}\hat{\vec{p}}\ket{a_1} = p_0 \cdot \hat{y}
\label{eq:MOdipole}
\end{equation}
where $\ket{a_1}$, $\ket{e_x}$ and $\ket{e_y}$ are molecule orbits of NV centers~\cite{Doherty2011}, $\hat{x}$ and $\hat{y}$ are unit vector pointing along $x$ or $y$ direction. We note that the state $\ket{e_y}$ has intrinsic dipole moment and $\bra{e_x}\hat{\vec{p}}\ket{e_y}$ is non-zero. However, since we only consider the transition between spin-triplet ground states and excited states of an NV center, the assumption in Eq.~\eqref{eq:MOdipole} is enough. The equality of the magnitude of these two dipole moment is guaranteed by Wigner-Echart theorem.

Using Eq.~\eqref{eq:MOdipole} with Table 1 (and Table A.1) in Ref.~\cite{Doherty2011}, we can calculate the dipole moment operators between the electronic fine levels of ground and excited states. Here we only consider spin $1$ states whose energy is inside the diamond band gap. Because the dipole transition does not interact with spin degree of freedom, the spin projection along $z$ direction should be invariant. The non-zero dipole moment operator elements between definite orbital symmetry states are:
\begin{equation}
\begin{aligned}
\bra{A_2,1,0} &\hat{\vec{p}} \ket{E_x,1,0} & = p_0 \cdot \hat{y} 
\\ \bra{A_2,1,0} &\hat{\vec{p}} \ket{E_y,1,0} & = p_0 \cdot \hat{x} 
\\ \bra{A_2,1,+1} &\hat{\vec{p}} \ket{E_x,1,+1} & = p_0 \cdot \hat{y} 
\\ \bra{A_2,1,+1} &\hat{\vec{p}} \ket{E_x,1,+1} & = p_0 \cdot \hat{x} 
\\ \bra{A_2,1,-1} &\hat{\vec{p}} \ket{E_x,1,-1} & = p_0 \cdot \hat{y} 
\\ \bra{A_2,1,-1} &\hat{\vec{p}} \ket{E_x,1,-1} & = p_0 \cdot \hat{x} 
\end{aligned}
\end{equation}
Here the states are labeled as $\ket{k,S,S_z}$, where $k$ labels the lattice symmetry group irreducible representations, $S$ is the spin quantum number, $S_z$ is the $z$-direction spin projection quantum number.
These states can be found in Ref.~\cite{Doherty2011} Table~1 and Table~A.1. For completeness, we list them using hole representation here,
\begin{equation}
\begin{array}{ll}
\ket{A_2,1,0} & = \left( \ket{e_x \bar{e_y}} + \ket{\bar{e_x} e_y} \right)/ \sqrt{2} \\
\ket{E_x,1,0} & = \left( \ket{\bar{a_1} e_x} + \ket{a_1 \bar{e_x}} \right)/ \sqrt{2} \\
\ket{E_y,1,0} & = \left( \ket{\bar{a_1} e_y} + \ket{a_1 \bar{e_y}} \right)/ \sqrt{2} \\
\ket{A_2,1,1} & = \ket{\bar{e_x} \bar{e_y}} \\
\ket{E_x,1,1} & = \ket{\bar{a_1} \bar{e_x}} \\
\ket{E_y,1,1} & = \ket{\bar{a_1} \bar{e_y}} \\
\ket{A_2,1,-1} & = \ket{e_x e_y} \\
\ket{E_x,1,-1} & = \ket{a_1 e_x} \\
\ket{E_y,1,-1} & = \ket{a_1 e_y} 
\end{array}
\end{equation}
where the bar denotes spin-down.

Similarly, we can also find the dipole moment operators between definite spin-orbital symmetry states which are shown in Table 1 of Ref.~\cite{Doherty2011}. The states $\ket{g_1}$, $\ket{g_2}$ and $\ket{g_3}$ are used to label states $\Phi_{1,A_1}^{\text{SO}}$, $\Phi_{2,E,x}^{\text{SO}}$ and $\Phi_{2,E,y}^{\text{SO}}$ in Ref.~\cite{Doherty2011} respectively. Since these states do not mix under spin-orbit and spin-spin interactions, we write them down explicitly here for ease of use later,
\begin{equation}
\begin{aligned}
\ket{g_1} = & \ket{A_2,1,0} \\
\ket{g_2} = & \frac{-1}{\sqrt 2} \left( \ket{A_2,1,1} - \ket{A_2,1,-1} \right) \\
\ket{g_3} = & \frac{-i}{\sqrt 2} \left( \ket{A_2,1,1} + \ket{A_2,1,-1} \right).
\end{aligned}
\label{eq:groundState}
\end{equation}
We also write down the excited fine levels with definite spin-orbit symmetry, which we label $\ket{e_1}$ to $\ket{e_6}$ here (these are labeled
$\Phi_{5,E,x}^{\text{SO}}$, $\Phi_{5,E,y}^{\text{SO}}$, $\Phi_{6,E,x}^{\text{SO}}$, $\Phi_{6,E,x}^{\text{SO}}$, $\Phi_{7,A_2}^{\text{SO}}$ and $\Phi_{8,A_1}^{\text{SO}}$ in Ref.~\cite{Doherty2011}):
\begin{equation}
\begin{array}{lll}
\ket{e_1} &=\Phi_{5,E,x}^{\textrm{SO}}=&\frac{1}{2} [ -i\left( \ket{E_x,1,1}+\ket{E_x,1,-1} \right) \\
 & & -\left( -\ket{E_y,1,1} + \ket{E_y,1,-1} \right)] \\
\ket{e_2} &=\Phi_{5,E,y}^{\textrm{SO}}=&\frac{1}{2} [- \left( -\ket{E_x,1,1}+\ket{E_x,1,-1} \right) \\\
 & & +i \left( \ket{E_y,1,1} + \ket{E_y,1,-1} \right)] \\
\ket{e_3} &=\Phi_{6,E,x}^{\textrm{SO}}=&-\ket{E_y,1,0} \\
\ket{e_4} &=\Phi_{6,E,y}^{\textrm{SO}}=&\ket{E_x,1,0} \\
\ket{e_5} &=\Phi_{7,A_2}^{\textrm{SO}}=&\frac{1}{2} [ \left( -\ket{E_x,1,1}+\ket{E_x,1,-1} \right) \\
& & +i \left( \ket{E_y,1,1} + \ket{E_y,1,-1} \right)] \\
\ket{e_6} &=\Phi_{8,A_1}^{\textrm{SO}}=&\frac{1}{2} [ -i\left( \ket{E_x,1,1}+\ket{E_x,1,-1} \right) \\
& & +\left( -\ket{E_y,1,1} + \ket{E_y,1,-1} \right)]
\end{array}
\end{equation}
The non-zero dipole moment operator matrix elements can be calculated for states of definite spin-orbital (SO) symmetry using the molecular orbitals. The dipole moment operators between the SO ground and excited state are labeled $\hat{\vec{p}}_{i,j}=\bra{g_i} \hat{\vec{p}} \ket{e_j}$, and can be represented as a matrix:
\begin{equation}
\hat{\vec{p}}_{i,j} = p_0 \cdot
\left( \begin{array}{cccccc}
\bm{0} & \bm{0} & \hat{x} & \hat{y} & \bm{0} & \bm{0}\\
-\frac{\hat{x}}{\sqrt 2} & \frac{\hat{y}}{\sqrt 2} & \bm{0} & \bm{0} & -\frac{\hat{y}}{\sqrt 2} & \frac{\hat{x}}{\sqrt{2}}\\
\frac{\hat{y}}{\sqrt 2} & \frac{\hat{x}}{\sqrt 2} & \bm{0} & \bm{0} & \frac{\hat{x}}{\sqrt 2} & \frac{\hat{y}}{\sqrt{2}}
\end{array}
\right).
\label{eq:SOStates}
\end{equation}
Here $\bm{0}$ indicates forbidden in dipole transitions. Note, this dipole moment operator matrix is consistent with the group symmetry prediction shown in Table A.4 of the Ref.~\cite{Doherty2011}. 

Furthermore, the spin-orbit interaction and spin-spin (SS) Hamiltonian given in the basis of SO states can be found in Ref.~\citep{Doherty2011} Table 2 and Table 3. Due to the large energy separation between the electronic ground states and excited states, the matrix elements out of the block of ground states or excited states are ignored, i.e. the perturbation theory can applied to the electronic ground states and excited states separately. The perturbation Hamiltonian for SO and SS interactions in ground state manifold, $V_{g} = V_{g}^{\text{(SO)}} + V_{g}^{\text{(SS)}}$, is diagonal, which means the states $\ket{g_1}$, $\ket{g_2}$ and $\ket{g_3}$ are still the eigenstates of the NV-center with SO interaction ($V_{g}^{\text{(SO)}}$) and SS interaction ($V_{g}^{\text{(SS)}}$). However, the perturbation Hamiltonian in the excited state manifold, $V_{e} = V_{e}^{(SO)} + V_{e}^{(SS)}$, is not diagonal. Besides affecting the level splitting, the perturbation interaction Hamiltonian results in mixing of the excited state. 

We can find a unitary matrix $U_e$ to diagonalize the excited state perturbation Hamiltonian $V_{e}$ by $U_{e} V_{e} U_{e}^{\dagger}$. The eigenstates of the new basis can be transformed from the SO basis by applying the unitary matrix $U_e$ to the SO basis. Therefore, the dipole moment operator between the ground states and the new excited states can be found by treating $\left( \hat{\vec{p}}_{i,j} \right)$ in Eq.~\eqref{eq:SOStates} as a matrix and applying $\left( \hat{\vec{p}}_{i,j} \right) \cdotp U_{e}^{\dagger}$. After taking the SS interactions into consideration, the excited state $\ket{e_1}$ mixes with  state $\ket{e_3}$, state $\ket{e_2}$ mixes $\ket{e_4}$, which results in small but non-zero dipole moment matrix elements between ground states $\ket{g_2}$ and $\ket{g_3}$ to the excited states $\ket{e_3}$ and $\ket{e_4}$. The eigenstates that diagonalize the SO and SS interaction Hamiltonian in NV electronic excited states are noted as SS basis of the NV center excited states and they are labeled as $\ket{\tilde{e}_i}$ for $i=1$ to $6$. Note that the notation $\ket{e_i}$ in our main paper refers to the SS basis states instead. The dipole moment operator between NV ground states and SS basis states of excited states is
\begin{equation}
\frac{\hat{\vec{p}}}{p_0}=
\left( \begin{array}{cccccc}
-F_{11} \hat{x} & -F_{11} \hat{y} & F_{12} \hat{x} & F_{12} \hat{y} & \vec{0} & \vec{0} \\
-F_{21} \hat{x} & F_{21} \hat{y} & -F_{22} \hat{x} & F_{22} \hat{y} & -F_{23} \hat{y} & F_{23} \hat{x} \\
F_{21} \hat{y} & F_{21} \hat{x} & F_{22} \hat{y} & F_{22} \hat{x} & F_{23} \hat{x} & F_{23} \hat{y} 
\end{array}
\right)
\label{eq:dipole_SS}
\end{equation}
where $F_{11} = 0.0513 $, $F_{12} = 0.9987$, $F_{21}=0.7062$, $F_{22} = 0.0363$, $F_{23}=1/\sqrt{2}$.

The strain field ($\vec{\xi}$) can also affect the NV electronic states. The strain field interactions to the NV electronic ground states are much smaller than the interactions to the excited states. Therefore we ignore the strain interaction to the NV ground states and only consider the excited state mixing due to the strain field. According to Ref.~\cite{Doherty2011}, axial strain field ($\xi_z$) does not mix the excited states, it only shifts the energy of the excited states and hence the dipole moment matrix does not change. However, the interaction Hamiltonian due to transverse strain field $\xi_x$ and $\xi_y$ has off-diagonal matrix elements in the SO basis of excited states, which means the transverse strain field mixes the SO basis of excited states.

Assume the transverse strain field is small so that the group symmetry of NV center is still preserved. The interaction Hamiltonian for $\hat{x}$-direction strain field is
\begin{equation}
H(\xi_x) =
\left( \begin{array}{cccccc}
0 & 0 & 0 & 0 & 0 & -E \\
0 & 0 & 0 & 0 & E & 0\\
0 & 0 & E & 0 & 0 & 0\\
0 & 0 & 0 & -E & 0 & 0\\
0 & E & 0 & 0 & 0 & 0 \\
-E & 0 & 0 & 0 & 0 & 0
\end{array}
\right)
\label{eq:x_strain}
\end{equation}
in the basis of the SO basis states, where $E$ is the interaction strength introduced by $\hat{x}$ direction strain field. From the Hamiltonian, the excited state $\ket{e_1}$ mixes with state $\ket{e_6}$, state $\ket{e_2}$ mixes with state $\ket{e_5}$. Since the dipole moment between the states $\ket{e_1}$, $\ket{e_6}$ and ground states has the same direction, we should expected that the dipole moment elements between SS basis states $\bra{\tilde{e}_1} \hat{\vec{p}} \ket{g_j}$ and $\bra{\tilde{e}_6} \hat{\vec{p}} \ket{g_j}$ for $j=2,3$ does not change directions, which can be easily checked after diagonalize the SO, SS with the strain field coupling Hamiltonian. Similar to the $\bra{\tilde{e}_2} \hat{\vec{p}} \ket{g_j}$ and $\bra{\tilde{e}_5} \hat{\vec{p}} \ket{g_j}$. Besides, due to the perturbation introduced by $\hat{x}$-direction strain field, the degeneracy of excited states $\ket{\tilde{e}_1}$ and $\ket{\tilde{e}_2}$ as well as the degeneracy of states $\ket{\tilde{e}_3}$ and $\ket{\tilde{e}_4}$ is broken.

The Hamiltonian for small $\hat{y}$-direction strain field in diamond crystal is,
\begin{equation}
H(\xi_y) =
\left( \begin{array}{cccccc}
0 & 0 & 0 & 0 & -E & 0\\
0 & 0 & 0 & 0 & 0 & -E \\
0 & 0 & 0 & -E & 0 & 0\\
0 & 0 & -E & 0 & 0 & 0\\
-E & 0 & 0 & 0 & 0 & 0 \\
0 & -E & 0 & 0 & 0 & 0
\end{array}
\right)
\label{eq:y_strain}
\end{equation}
where $E$ is the interaction energy due to the $\hat{y}$ direction strain field. The $\hat{y}$ direction strain field mixes the excited state $\ket{e_1}$ with $\ket{e_5}$, state $\ket{e_2}$ with $\ket{e_6}$ and state $\ket{e_3}$ with $\ket{e_4}$. The dipole moment $\bra{\tilde{e}_i} \hat{\vec{p}} \ket{g_j}$ for $i=1$ to $6$ and $j=2,3$ does not point along $\hat{x}$ or $\hat{y}$ directions any more. Instead, the dipole moment between the same excited state and the two ground states $\ket{g_2}$ and $\ket{g_3}$ are no longer orthogonal. This feature of the dipole moment matrix causes that the scattering light from state-preserving and state-flipping transitions are not polarized along perpendicular directions.

\subsection{Transition rates and Raman photon polarization} \label{subsec:transition_rates}
In this section, we present the details of the scattering rate calculation. To estimate the magnitude of the dipole moment, we modeled the relaxation from the electronic excited state with $S_z=0$ (e.g $\ket{e_3}$), back to ground state with $S_z=0$ (e.g. $\ket{g_1}$) as a two-level system spontaneous relaxation process. If we ignore the slow relaxation processes from state $\ket{e_3}$ to the other two ground state levels $\ket{g_2}$ and $\ket{g_3}$, then the lifetime of state $\ket{e_3}$, which is  $13$~ns~\cite{Doherty2013}, can be used to estimate the value of dipole moment. The magnitude of dipole moment estimated based on this method is $\lvert p \rvert = e \lvert d \rvert = 5.2$ Debye~\cite{Alkauskas2014}, where $e$ is the electron charge.

As we pointed out in~\ref{subsec:waveguide} and~\ref{subsec:dipole} the NV center dipole moments for optical transition between ground and excited states are along the transverse direction. Therefore, we choose to match the axial direction of NV centers ($\hat{z}$ direction) to the waveguide $\hat{z}$ direction to have optimum coupling efficiency. We also choose to match the NV center intrinsic transverse directions $\hat{x}$ and $\hat{y}$ with the waveguide transverse direction $\hat{x}$ and $\hat{y}$ as Fig.~\ref{fig:MBEphonon}(c) shows.

To calculate the scattering transition rates between ground states $\ket{g_2}$ and $\ket{g_3}$, we consider a single NV center residing inside an infinitely long waveguide shown in~\ref{subsec:waveguide}. The quantized guided waveguide mode in a length $L$ waveguide, with wavevector along the waveguide axial direction $k_z$ and mode index $m$ is~\cite{Lodahl2015}:
\begin{equation}
\hat{E}_{k_z,m} = \mathcal{E}_0 (k_z) \vec{u}_{k_z,m} (x,y) a_{k_z,m} \frac{1}{\sqrt{L}} e^{i k_z z - i \omega_{k_z}t} + h.c.,
\label{eq:quantize_guide_mode}
\end{equation}
where $a_{k_z,m}$ is the annihilation operator for photons with $k_z$ and mode $m$, $\omega_{k_z}$ is the angular frequency of the mode photon, which can be determined by the waveguide dispersion relations, $\mathcal{E}_0 (k_z) = \sqrt{\hbar \omega_{k_z}/2 \varepsilon_0}$ in which $\varepsilon_0$ is the vacuum permittivity,  $\vec{u}_{k_z,m}(x,y)$ is the mode profile on the cross section of the waveguide. The mode profile is normalized according to the normalization condition,
\begin{equation}
\int dx dy \, \varepsilon_r (x,y) \vec{u}_{k_z,m}^{*} (x,y) \cdot \vec{u}_{k_z,n} (x,y) = \delta_{m,n}
\label{eq:mode_normalization_2}
\end{equation}

To simplify the calculation, we assume the NV centers only couple to the driving light and the waveguide modes, and ignore the coupling to the non-guided modes. We further assume the driving light is a classical field while the waveguide modes are quantized. The interaction Hamiltonian is,
\begin{widetext}
\begin{equation}
\begin{aligned}
H_{\text{int}} = & H_{\text{drive}} + H_{\text{guide}} \\
H_{\text{drive}} = & \left[ \sum_{i,j} \vec{E}_{d}^{*}(\vec{r}_0) \cdot \hat{\vec{p}}_{i,j} \ket{g_i}\bra{e_j} e^{i \left( \omega_d - \omega_{ej,gi}\right) t} +h.c. \right]\\
H_{\text{guide}} = & \left[ \sum_{i,j} \sum_{k_z} \sum_{m_k} \mathcal{E}_0(k_z) \left( \vec{u}_{k_z,m_k} (\vec{r}_0) \cdot \hat{\vec{p}}_{i,j}^{*} \right) a_{k_z,m} \ket{e_j}\bra{g_i} e^{i \left( \omega_{ej,gi}-\nu_{\vec{k},\lambda}\right)}+ h.c.\right].
\end{aligned}
\label{eq:RamanRate}
\end{equation}
\end{widetext}
$H_{\text{drive}}$ is for the interaction between the NV center and the driving light. The classical electromagnetic field, $\vec{E}(\vec{r}) e^{i \omega_d t}$, is the driving laser light. $\hat{\vec{p}}_{i,j}$ is defined as $\bra{g_i} \hat{\vec{p}} \ket{e_j}$, where $\ket{e_j}$ is the eigenstates of electronic excited state of NV center. $H_{\text{guide}}$ is for the interaction with the waveguide guided modes, $\vec{r}_0$ is the position of the NV center. The summation index $i = 1$ to $3$, while index $j=1$ to $6$. The mode index $m$ goes through all the guided modes in the waveguide with wave vector $k_z$.

Note that the photon scattering process from ground state $\ket{g_i}$ to the ground state $\ket{g_{i'}}$ is a second order process. We use second order Fermi's golden rule to calculate the transition rates. Assuming that initially there are no photons in the guided modes, and hence the initial state is $\ket{\Psi_i}=\ket{g_i}\otimes \ket{0}$, where $\ket{0}$ is the vacuum guided mode fields, while the scattering final state is $\ket{\Psi_f}=\ket{g_{i'}}\otimes\ket{1_m}$, where $\ket{1_m}$ is the state for one photon inside the guided mode $m$. Based on the second order Fermi's Golden Rule, the transition rate from initial state $\ket{g_i}\otimes \ket{0}$ to final state $\ket{g_{i'}}\otimes \ket{1_m}$ is,
\begin{equation}
\begin{aligned}
& \Gamma_{i \rightarrow i'} = \frac{2 \pi}{\hbar} \delta(\epsilon_{f}-\epsilon_{i})  \\
& \times \left\vert \sum_{j=1}^{6} \bra{\Psi_f} \frac{H_{\text{guide}}\ket{e_j}\ket{0}\bra{0}\bra{e_j}H_{\text{drive}}}{\hbar \omega_d + \epsilon_{g,i}- \epsilon_{e,j}} \ket{\Psi_i}\right\vert^2
\end{aligned}
\end{equation}
where $\epsilon_{g,i}$ and $\epsilon_{e,j}$ are for the energy of NV states $\ket{g_i}$ and $\ket{e_j}$, $\omega_d$ is the driving light angular frequency. We define an effective Hamiltonian for Raman transition as,
\begin{equation}
\begin{aligned}
& H_{\textrm{eff}} = \sum_{j=1}^{6} \frac{H_{\text{guide}}\ket{e_j}\ket{0}\bra{0}\bra{e_j}H_{\text{drive}}}{\hbar \omega_d + \epsilon_{g,i}- \epsilon_{e,j}} \\
& = \sum_{k_z,m} \sum_{j=1}^{6} \frac{\mathcal{A}_{k_z,m}(\vec{r}_0)}{\Delta_j}  \left( \hat{u}_{k_z,m}\cdot \hat{p}_{i',j}\right)^{*}\left( \hat{\lambda}_d \cdot \hat{p}_{i,j}\right) a_{k_z,m}^{\dagger}
\end{aligned}
\label{eq:eff_H}
\end{equation}
where $\mathcal{A}_{k_z,m}(\vec{r}_0)$ is a constant defined as $\mathcal{E}_0(k_z) u^{*}_{k_z,m} E_d p_{0}^2$, energy mismatch $\Delta_j$ is defined as $\hbar \omega_d + \epsilon_{g,i} - \epsilon_{e,j}$. The variable $u_{k_z,m}$ is the magnitude of the waveguide mode with wave-vector $k_z$ and mode index $m$ at the NV position $\vec{r}_0$, $\hat{u}_{k_z,m}$ is the unit vector along the electric field of the mode at the NV center location, $\hat{p}_{i,j}$ is defined as $\hat{p}_{i,j} = \vec{p}_{i,j} / p_0$ in which $\vec{p}_{i,j}$ is the dipole moment operator elements between ground state $\ket{g_i}$ and excited $\ket{e_j}$. The driving field magnitude at the NV location is noted as $E_d$, while its polarization direction is labeled as $\hat{\lambda}_d$. The transition amplitude can be written as $\bra{\Psi_f} H_{\textbf{eff}} \ket{\Psi_i}$. 

As we pointed out in~\ref{subsec:waveguide}, at the ``magic'' frequency, there are only two guided modes supported by the diamond waveguide. Further, mode $1$ and mode $2$ only have non-zero $E_x$ or $E_y$ components respectively (when the NV center is centered in the waveguide: $x,y\sim 0$). Therefore, the transitions with $\hat{x}$ dipole and transitions with $\hat{y}$ dipole couple to different modes. If we also assume that at the NV center location, $E_x(\vec{r}_0)$ of mode $1$ is equal to $E_y(\vec{r}_0)$ of mode $2$, the constant $\mathcal{A}$ does not depend on mode number $m$. If we only considered the modes which respect the energy conservation, and use $\hat{x}$ polarized light to drive the transitions, the effective Hamiltonian can be written as,
\begin{widetext}
\begin{equation}
\begin{aligned}
H_{\textrm{eff},k_{z0}}/\mathcal{A}_{k_{z0}} & =
\left( \frac{F_{21}^2}{\Delta_1} + \frac{F_{22}^2}{\Delta_3} + \frac{F_{23}^2}{\Delta_6}\right)\ket{g_2}\bra{g_2} a_{k_{z0},2}^{\dagger}
+\left( \frac{F_{21}^2}{\Delta_2} + \frac{F_{22}^2}{\Delta_4} + \frac{F_{23}^2}{\Delta_5}\right)\ket{g_3}\bra{g_3} a_{k_{z0},2}^{\dagger} \\ 
& + \left( \frac{-F_{21}^2}{\Delta_1} + \frac{-F_{22}^2}{\Delta_3} + \frac{F_{23}^2}{\Delta_6}\right)\ket{g_3}\bra{g_2} a_{k_{z0},1}^{\dagger} 
 + \left( \frac{F_{21}^2}{\Delta_2} + \frac{F_{22}^2}{\Delta_4} + \frac{-F_{23}^2}{\Delta_5}\right)\ket{g_2}\bra{g_3} a_{k_{z0},1}^{\dagger}
\end{aligned}
\label{eq:detail_eff_H}
\end{equation}
\end{widetext}
where we adopt the dipole moment operator expression in Eq.~\eqref{eq:dipole_SS}. The first and second terms give the state-preserving transitions, while the third and fourth terms give the state-flipping transitions. According to Eq.~\eqref{eq:detail_eff_H}, photons from state-preserving transitions and state-flipping transitions have perpendicular polarizations, and hence they couple to two different modes. Similarly, if the driving light is polarized along $\hat{y}$ direction, following the same argument, it is easy to show that the photons from state-preserving transitions are coupled to mode $2$, while photons from state-preserving transitions are coupled to the mode $1$ instead. The orthogonal polarization of photons is a feature that originates in the orthogonal dipole moment between the ground states $\ket{g_2}$, $\ket{g_3}$ and the same excited state $\ket{e_j}$, i.e.
\begin{equation}
\bra{g_2}\hat{\vec{p}}\ket{e_{j}} \cdot \bra{g_3} \hat{\vec{p}} \ket{e_{j}} = 0 
\label{eq:dipole_orthogonal}
\end{equation}
for $j=1$ to $6$ (we call this property orthogonality). The perturbation on the excited state energy, the dipole moment elements and the $\hat{x}$ direction strain field interaction, does not change this dipole moment property, and hence  orthogonal polarization of photons is still expected from state-preserving and state-flipping transitions. If this feature does not persist, e.g. adding $\hat{y}$ direction strain field, the photons coming from state-flipping and state-preserving transitions become non-orthogonally polarized.

The ``magic'' point is the point where both state-preserving transitions are highly suppressed. According to the Eq.~\eqref{eq:detail_eff_H}, this requires,
\begin{equation}
\begin{aligned}
& \frac{F_{21}^2}{\Delta_1} + \frac{F_{22}^2}{\Delta_3} + \frac{F_{23}^2}{\Delta_6} =0 \\
& \frac{F_{21}^2}{\Delta_2} + \frac{F_{22}^2}{\Delta_4} + \frac{F_{23}^2}{\Delta_5} =0 
\end{aligned}
\end{equation}
However, there is no driving light frequency that can satisfy both equations. Instead, we choose to minimize the larger rates of these two transitions to improve the gate fidelity, i.e. to minimize $$\text{Max}\left[ \left\vert \frac{F_{21}^2}{\Delta_1} + \frac{F_{22}^2}{\Delta_3} + \frac{F_{23}^2}{\Delta_6}\right\vert, \, \left\vert\frac{F_{21}^2}{\Delta_2} + \frac{F_{22}^2}{\Delta_4} + \frac{F_{23}^2}{\Delta_5}\right\vert \right].$$ We found this is equivalent to solving the equation:
\begin{equation}
\left( \frac{F_{21}^2}{\Delta_1} + \frac{F_{22}^2}{\Delta_3} + \frac{F_{23}^2}{\Delta_6}\right)^2 = \left(\frac{F_{21}^2}{\Delta_2} + \frac{F_{22}^2}{\Delta_4} + \frac{F_{23}^2}{\Delta_5}\right)^2,
\end{equation}
which gives the frequency of the ``magic'' point used in the main manuscript.

The transition rates at the ``magic'' point can be calculated using Fermi's golden rule. We sum over all the possible $k_z$ and $m$ to get the transition rate from the initial state $\ket{g_i}$ to final state $\ket{g_{i'}}$:
\begin{equation}
\begin{aligned}
\Gamma_{i \rightarrow i'} = & \frac{\pi n_{\textrm{eff}} \omega_d p_{0}^{4} \vert u \vert^2 \vert E_d \vert^2}{c\hbar \varepsilon_0} \\
& \times \left\vert \sum_{j,m}\frac{1}{\Delta_j}\left( \hat{u}_{m}\cdot \hat{p}_{i',j}\right)^{*}\left( \hat{\lambda}_d \cdot \hat{p}_{i,j}\right) \right\vert^2.
\end{aligned}
\label{eq:rate_v1}
\end{equation}
Here, $n_{\textrm{eff}}$ is the effective refractive index for the modes at the frequency of the driving light, the dispersion relation of the guided modes at the driving light frequency is $\omega = (c/n_{\textrm{eff}}) k_z$. We also assume the NV center is located at a point where the $E_x$ field of mode $1$ is equal to the $E_y$ field of mode $2$, which is represented as $u$, while the $E_y$ of mode $1$ and $E_x$ of mode $2$ is zero. The unit vectors $\hat{u}_m$ and $\hat{\lambda}_d$ shows the direction of the guided field in waveguide and the driving field at the NV location. To convert the term inside $|\dots|^2$ to a dimensionless parameter, we define $ \Delta_{j} = h \nu_0 \tilde{\Delta}_j$ where $\nu_0 = 1$~GHz. Therefore we can define a rate constant $\Gamma_0$ and a dimensionless parameter $\mathcal{G}_{i,i'}$ so that the transition rate $\Gamma_{i \rightarrow i'} = \Gamma_0 \mathcal{G}_{i,i'}(\omega_d)$, where 
\begin{align}
\Gamma_0 & = \frac{n_{\textrm{eff}} \omega_d p_{0}^{4} \vert u \vert^2 \vert E_d \vert^2}{4\pi c\hbar^3 \varepsilon_0 \nu_0} \\
\mathcal{G}_{i,i'} & = \left\vert \sum_{j,m}\frac{1}{\tilde{\Delta}_j} \left( \hat{u}_{m}\cdot \hat{p}_{i',j}\right)^{*}\left( \hat{\lambda}_d \cdot \hat{p}_{i,j}\right) \right\vert^2
\end{align}

\begin{figure}[htbp!]
\includegraphics[width = 3.2 in]{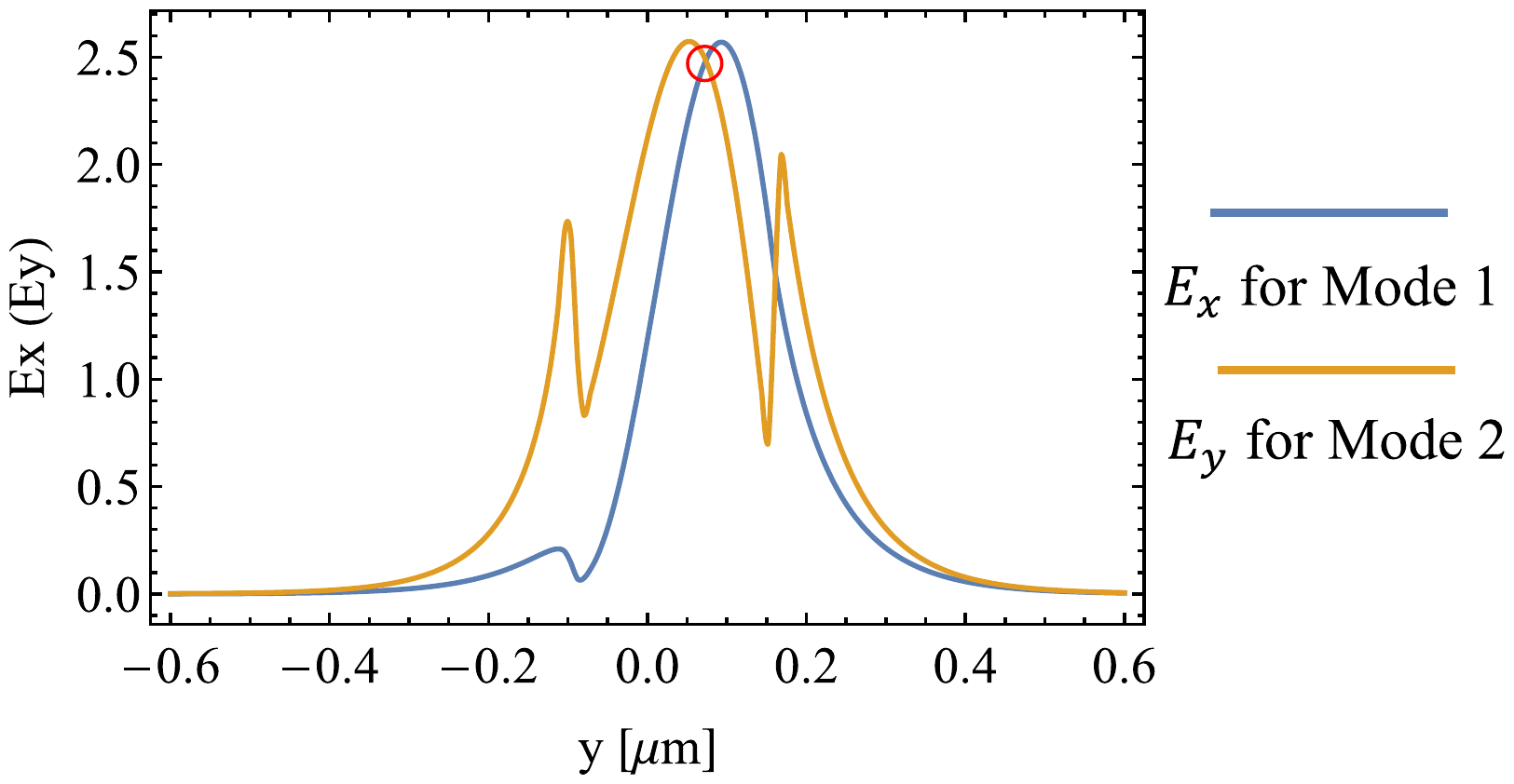}
\caption{$E_x$ component of mode $1$ and $E_y$ component of mode $2$ at $x=0$ of the waveguide. We can find a point (red circle) that satisfies $E_{x,1} = E_{y,2}$.}
\label{fig:field_slice}
\end{figure}

By solving the mode profiles at the ``magic'' frequency, the effective refractive index of these two modes are $n_{\textrm{eff}}=1.580$. At $x=0$, after properly normalize the mode fields using Eq.~\eqref{eq:mode_normalization_2}, we can find a point which satisfies our assumptions, i.e. $E_{x,1}(y_0) = E_{y,2}(y_0)$ (see Fig.~\ref{fig:field_slice}). At this point, $u = 2.4847 \, \mu\textrm{m}^{-1}$. We estimate the electric field of the driving light by a $1\, \mu\textrm{W}$ plane wave focused with a $1\, \mu\textrm{m}^2$ region. The  transition rate constant is calculated as $\Gamma_0 = 20.78$~MHz.

\subsection{Gate fidelity and tolerance of the magic point against NV electronic state perturbation}
In this section, we provide a more detailed discussion and analysis of how perturbations to NV electronic states affect the drive frequency (especially the ``magic'' frequency) and the gate fidelity. We focused on three types of perturbations: (1) shifts of the excited state energy /effect
of an NV center, (2) perturbation of the dipole moment matrix elements and (3) small transverse strain fields inside the diamond crystal. We also analyze how each of the perturbation affects the polarization of the emitted photons. We mainly focus on the effect of perturbation at the ``magic'' point and explore how these perturbations affect gate fidelity for the gate operation schemes $M1$, $M2$, and $M3$. 

\begin{figure*}[!htbp]
\includegraphics[width=\textwidth]{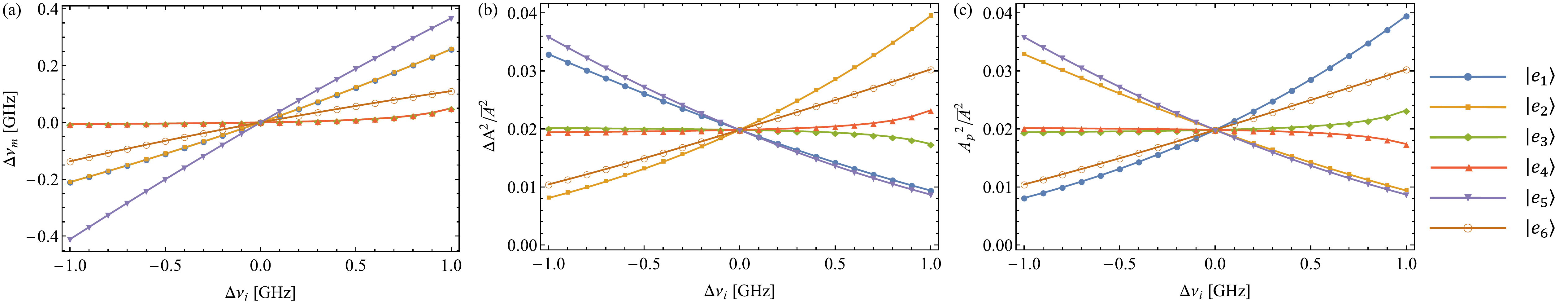}
\caption{Tolerance of the ``magic'' point to shifts of the fine levels of the NV excited states. We perturb the energy of each level ($\Delta E_{e,i}$) by $\pm 1$ GHz. In (a), we plot the shift of the ``magic'' frequency relative to the unperturbed case as we perturb the energy of each excited state. Shifting the energy of excited states also affects the state-preserving and state-flipping transition rates. In (b), we plot the gate infidelity of scheme $M1$ due to the imbalance in the state-flipping transition rates as we perturb the energy of each excited state. In (c), we plot the gate infidelity of scheme $M2$ caused by the leakage of the state-preserving photons as we perturb the energy of each excited state.}
\label{fig:energy_shift}
\end{figure*}

First, we consider perturbations that shift the energy of NV excited states. Since this type of perturbations does not affect the dipole moment between the ground states and excited states, the orthogonal property of scattered photon polarizations that are utilized by $M1$ and $B1$ are preserved. However, shifts of the excited state energies changes the transition amplitudes and hence may shift the position of the ``magic'' point. Changes in the state-flipping amplitudes affect the imbalance of the two state-flipping transitions rates, thus affect gate fidelity in scheme $M1$. Changes of the state-preserving transition amplitudes affect the suppression at the ``magic'' frequency, which affects the gate fidelity of scheme $M2$. 

To quantitatively explore the effects of the shifting of NV center electronic excited states, we artificially shift the energy of the excited states $\ket{e_1}$ to $\ket{e_6}$ one-by-one by $\pm 1$~GHz, while leave the dipole moments unchanged.  With the energy level perturbation, we search around the original ``magic'' frequency to find a new ``magic'' frequency that minimize both state-flipping transition amplitudes. The shift of the ``magic'' frequency as we shift each of the excited state energies is plotted in Fig.~\ref{fig:energy_shift}(a). 

Assuming that the imbalance of the two state-flipping transition amplitudes is small, i.e. $\frac{\vert A_1 - A_2 \vert}{A_1 + A_2} \ll 1$, where $A_1$ and $A_2$ are defined in Eq.~\eqref{eq:trans_amp_mgcpt}, enables us to expand the gate fidelity of scheme $M1$ as:
\begin{equation}
F_{e,1} = \frac{\left(A_1+A_2\right)^2}{2\left( A_1^2 + A_2^2 \right)} = \frac{\bar{A}^2}{\bar{A}^2 + \Delta A^2} \sim 1 - \frac{\Delta A^2}{\bar{A}^2}
\label{eq:m1_fid}
\end{equation}
where $\bar{A} = (A_1+A_2)/2$ and $\Delta A = \vert A_1 - A_2 \vert/2$. We calculate the gate infidelity ($1-F_{e1}$) in each cases with gate operation scheme $M1$ and show it in Fig.~\ref{fig:energy_shift}(b). As we shift each excited state energy of the NV center by $\pm 1$~GHz, the gate fidelity of gate operation scheme $M1$ is only slightly affected. In the worst case, when we shift the energy of state $\ket{e_2}$ by $+1$~GHz, the gate fidelity drops to $\sim 0.96$. 

The gate operation scheme $M2$ is not affected by the imbalance of state-flipping transitions. However, because the state-preserving transition relation $\frac{A^{x}_{p,2}}{A_{0}^{(x)}} = \frac{A^{y}_{p,3}}{A_{0}^{(y)}}=-\frac{A^{y}_{p,2}}{A_{0}^{(y)}} = -\frac{A^{x}_{p,3}}{A_{0}^{(x)}}$ holds, when drive light is polarized along $(\hat{x}+\hat{y})$ direction, the state-preserving scattered photons are still along $(\hat{x}-\hat{y})$ direction, which causes leakage of the state-preserving photons to the detector. Since we are working at the ``magic'' point where the state-preserving transitions are highly suppressed, we can also expand the gate fidelity of gate operation scheme $M2$ as:
\begin{equation}
F_{e,2} = \frac{\bar{A}^2}{\bar{A}^2 + A_{p}^2} \sim 1 - \frac{A_{p}^2}{\bar{A}^2}
\label{eq:m2_fid}
\end{equation}
where $A_{p}$ is the magnitude of the state-preserving transition amplitudes. In Fig.~\ref{fig:energy_shift}(c), we plot the gate infidelity of the scheme $M2$. When shifting energy of state $\ket{e_1}$ by $+1$~GHz, the gate infidelity increases $\sim 0.04$. Again, the gate operation fidelity is only slightly affected by the excited state energy level shifting.

Scheme $M3$ is not effected by shifting the excited state levels. Because the dipole moment is not affected, when the drive light is polarized along $(\hat{x} + \hat{y})$ direction, the state-preserving photons are still polarized along $(\hat{x}-\hat{y})$ direction. The collection path polarizer along $(\hat{x} + \hat{y})$ can fully eliminate the state-preserving photons. The polarizations of the two types of state-flipping photons still deviated from $(\hat{x}-\hat{y})$ direction by $\pm \theta$ (see Fig.~\ref{fig:polarization}), where $\theta$ is determined by the imbalance of the state-flipping transitions. However, since these two directions are centered on the direction $(\hat{x} - \hat{y})$, after the polarizer, the two state-flipping transition rates are balanced.

\begin{figure*}[!htbp]
\includegraphics[width = 5 in]{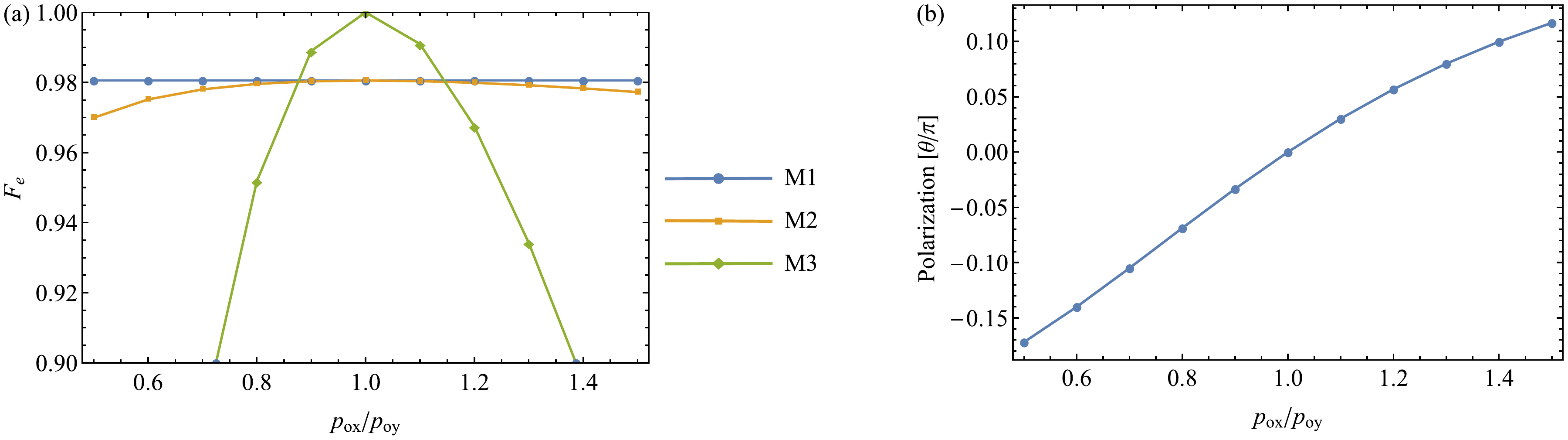}
\caption{We plot the entanglement gate fidelity for gate operation schemes $M1$ to $M3$ at the ``magic'' frequency of the drive light when we add dipole moment mismatch in (a). If the NV centers are driven by a $(\hat{x}+\hat{y})$ polarized light, because the four transition amplitudes are not all balanced, the two kinds of state-preserving photons, i.e. $\ket{g_2} \rightarrow \ket{g_2}$ and $\ket{g_3} \rightarrow \ket{g_3}$, are no longer polarized along $(\hat{x}-\hat{y})$ direction. We plot the polarization angle of the state-preserving photons with respect to $(\hat{x}-\hat{y})$ direction as a function of dipole mismatch in (b).}
\label{fig:dipole_perturb}
\end{figure*}

Second, we explore the effect of perturbations that modify the dipole moments of the NV centers. In~\ref{subsec:dipole}, we constructed the dipole moment using Eq.~\eqref{eq:MOdipole}. Let $\bra{e_x}\hat{\vec{p}}\ket{a_1} = p_{0x} \cdot \hat{x},  \bra{e_y}\hat{\vec{p}}\ket{a_1} = p_{0y} \cdot \hat{y}$, then $C_{3v}$ symmetry in combination with the Wigner-Eckart theorem guarantees that $p_{0x} = p_{0y}$, which is consistent with the assumptions in Eq.~\eqref{eq:MOdipole}. Here we assume there might be certain types of perturbations that break this relation and give $p_{0x}/p_{0y} \neq 1$. Notice, that these perturbations break the state-preserving amplitudes relation, i.e. $\vert \bra{g_2} \hat{\mathbf{p}} \ket{e_i} \vert \neq \vert \bra{g_3} \hat{\mathbf{p}} \ket{e_i} \vert$, which voids the origin of the equality of state-preserving transition amplitudes in Eq.~\eqref{eq:pres_amp_equality}. Therefore, we will have four different state-preserving transition amplitudes. If we assume $p_{0y} = p_0$, as we shift $p_{0x}$, in dipole moment matrix in Eq.~\eqref{eq:dipole_SS}, the components along $\hat{y}$ direction do not change, while the components along $\hat{x}$ change by a factor $O_{x} = p_{0x} / p_{0}$ and hence the state-preserving transition amplitudes become $\tilde{A}^{x}_{p,2} = O_{x}^{2} A^{x}_{p,2}$ and $\tilde{A}^{x}_{p,3} = O_{x}^{2} A^{x}_{p,3}$. 

At the unperturbed ``magic'' point, the state-preserving transition amplitudes satisfy $\frac{A^{x}_{p,2}}{A_{0}^{(x)}} = \frac{A^{y}_{p,3}}{A_{0}^{(y)}}=-\frac{A^{y}_{p,2}}{A_{0}^{(y)}} = -\frac{A^{x}_{p,3}}{A_{0}^{(x)}}$. Under the dipole moment perturbation we obtain:
\begin{equation}
\frac{\tilde{A}^{x}_{p,2}}{A_{0}^{(x)}} = -\frac{\tilde{A}^{x}_{p,3}}{A_{0}^{(x)}} = O_{x}^{2} \frac{\tilde{A}^{y}_{p,3}}{A_{0}^{(y)}} = - O_{x}^{2} \frac{\tilde{A}^{y}_{p,2}}{A_{0}^{(y)}}.
\label{eq:dipole_preserv_trans_amp}
\end{equation}
Even through we cannot suppress all four state-preserving transition amplitudes to the same level, we can still achieve a good suppression for $\tilde{A}^{x}_{p,2}$ and $\tilde{A}^{x}_{p,3}$ at the original ``magic'' point if the dipole mismatch factor $O_x$ is close to identity and hence we still use this drive frequency point as a ``magic'' point under perturbation.

We also notice that the orthogonality property of the dipole matrix persists, i.e. $\bra{g_2}\hat{\mathbf{p}}\ket{e_j} \cdot \bra{g_3} \hat{\mathbf{p}}\ket{e_j} = 0$ for $j=1$ to $6$. Due to this feature, if the drive is polarized along $\hat{x}$ or $\hat{y}$ direction, the state-flipping photons are polarized along the direction perpendicular to state-preserving photons. Hence, the drive and polarizer setup in $M1$ can fully eliminate the state-preserving Raman photons from the collection path. Moreover, according to the state-flipping transition amplitudes in Eq.~\eqref{eq:tran_amp_f}, when the perturbation gives mismatch factor $O_x \neq 1$, the state-flipping transition amplitudes are all enhanced (or shrunk) by a factor of $O_x$. Based on Eq.~\eqref{eq:m1_fid}, the gate fidelity for scheme $M1$ is not affected by the dipole moment perturbation, as shown in Fig.~\ref{fig:dipole_perturb}(a). 

When the drive is polarized along $(\hat{x} + \hat{y})$ direction, due to the fact that the four state-preserving transition amplitudes in Eq.~\eqref{eq:dipole_preserv_trans_amp} are not all equal at ``magic'' point, the state-preserving photons are not polarized along $(\hat{x}-\hat{y})$. We plot the deviation of the state-preserving transition photon polarization direction from $(\hat{x}-\hat{y})$ as the dipole mismatch changes in Fig.~\ref{fig:dipole_perturb}(b). Due to the rotation of the polarization direction of state-preserving photons, the state-preserving transition amplitudes seen after a $(\hat{x}-\hat{y})$ polarizer also varies. However, as the state-flipping transition amplitudes after the polarizer is much larger than the state-preserving transitions amplitudes, the gate operation scheme $M2$ is tolerant to small dipole mismatch as shown in Fig.~\ref{fig:dipole_perturb}(a). When the dipole moment mismatch is large (e.g. $\sim 0.5$), the gate fidelity of $M2$ drops by $\sim 0.01$. 

The gate fidelity of scheme $M3$ is strongly affected by the dipole moment perturbation as shown in Fig.~\ref{fig:dipole_perturb}(a). The polarizer setup in $M3$ is along $(\hat{x} + \hat{y})$ direction, which blocks most of the state-flipping photons. However, under the dipole moment perturbation, the state-preserving photons are not polarized along $(\hat{x}-\hat{y})$ direction, which breaks the unitarity of scheme $M3$. Further, the leakage of the state-preserving photons through the polarizer can be as strong as the state-flipping photons, which strongly affects the gate fidelity. Since the two kinds of state-preserving photons are linearly polarized along the same direction, it is possible to rotate the polarizer on the collection path to completely eliminate the state-preserving photons. However, the two state-flipping transitions seen after the polarizer are not balanced anymore. In this way, we can improve the fidelity of scheme $M3$, but the gate is no longer perfectly unitary.

\begin{figure*}[!htbp]
\includegraphics[width= \textwidth]{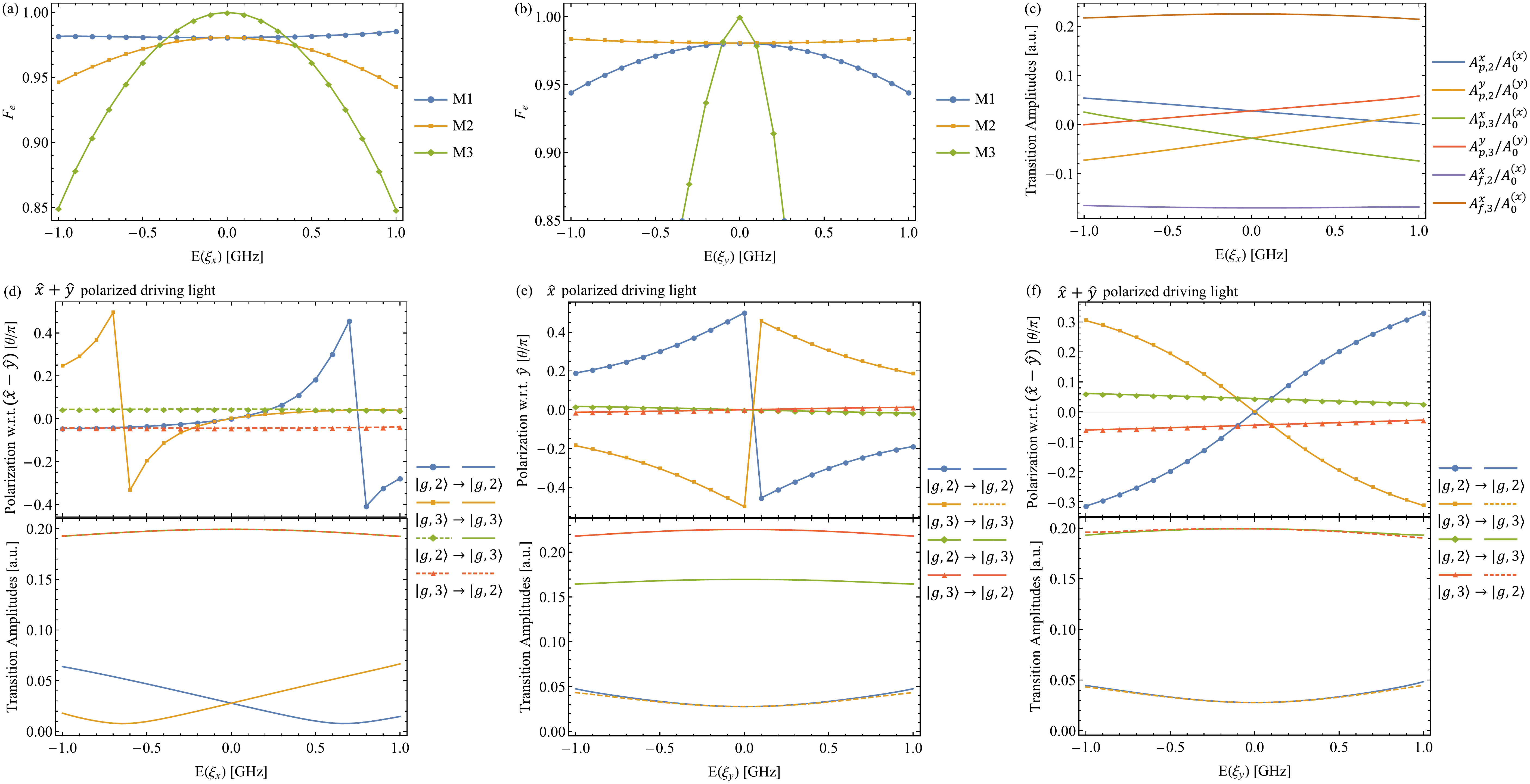}
\caption{We add $\hat{x}$ ($\hat{y}$) direction strain fields in diamond crystal to induce interaction with strength $E(\xi_x)$ [$E(\xi_y)$] [see Eq.\eqref{eq:x_strain} and Eq.\eqref{eq:y_strain}]. The gate entanglement fidelity of schemes $M1$, $M2$ and $M3$ are plotted in (a) for $\hat{x}$ direction strain field and (b) for $\hat{y}$ direction strain field. In (c), we apply $\hat{x}$ polarized driving light and plot the transition amplitudes as a function of the $\hat{x}$ direction strain field, $\xi_x$. We observe that the two state-flipping transition amplitudes are unbalanced and weakly affected by the strain. In (d), we apply $(\hat{x}+\hat{y})$ polarized driving light, and plot the polarization angles of state-preserving and state-flipping photons with respect to the $(\hat{x}-\hat{y})$ direction (top panel) and the magnitude of the state-preserving and state-flipping transition amplitudes (bottom panel) as a function of the $\hat{x}$ direction strain field. In (e), we apply $\hat{x}$ polarized driving light and plot the polarization angles of the state-preserving and state-flipping photons with respect to the $\hat{x}$ direction (top panel) and the magnitude of the state-preserving and state-flipping transition amplitudes (bottom panel) as a function of the $\hat{y}$ direction strain field. In (f), we apply $(\hat{x}+\hat{y})$ polarized driving light and plot the polarization angles of the state-preserving and state-flipping photons to the $(\hat{x}-\hat{y})$ direction (top panel) and the magnitude of the state-preserving and state-flipping transition amplitudes (bottom panel) as a function of the $\hat{y}$ direction strain field.}
\label{fig:strain_field}
\end{figure*}

Third, we consider perturbations due to a strain field in the diamond crystal. A strain field applied along the $\hat{x}$ ($\hat{y}$) direction mixes the NV excited states via the perturbation Hamiltonian Eq.~\eqref{eq:x_strain} (Eq.~\eqref{eq:y_strain}). The strain field also acts on the ground state manifold, however, it only shifts the energy of the $\ket{g_2}$ and $\ket{g_3}$ states. Here, we ignore the impact of the strain fields on the ground states and only focus on the excited states. Due to the mixing of the excited states, the dipole moment matrix does not preserve the property $\vert \bra{g_2} \hat{\mathbf{p}} \ket{e_i} \vert = \vert \bra{g_3}\hat{\mathbf{p}}\ket{e_i}\vert$ and hence we expect the four state-preserving transition amplitudes to be different. Moreover, in the presence of a strain field, it is impossible to find a frequency point to make all four transitions balanced. Instead, in the vicinity of the unperturbed ``magic'' frequency, there is a window of drive frequencies in which the state-preserving transitions are suppressed. Therefore, we can still use the unperturbed ``magic'' point as the drive frequency in the presence of a weak strain field.

Strain field applied in the $\hat{x}$ direction mixes the states $\ket{e_1}\leftrightarrow\ket{e_6}$, and $\ket{e_2} \leftrightarrow \ket{e_5}$. Note, the dipole moments between a certain ground state and the two excited states that are being mixed have the same direction. Hence, while the magnitude of the dipole moment between ground and excited states is affected by strain, its direction is not.  Therefore, the orthogonal properties of the dipole moment [see Eq.~\eqref{eq:dipole_orthogonal}] are preserved with the $\hat{x}$ direction strain field perturbation.

In Fig.~\ref{fig:strain_field}(a) we plot the gate entanglement fidelity for schemes $M1$, $M2$, and $M3$ as a function of strain in the $\hat{x}$ direction [expressed via the matrix element $E$ in Eq.~\eqref{eq:x_strain}]. We observe that strain has essentially no effect on the $M1$ scheme, weak effect on the $M2$ scheme, and strong effect on the $M3$ scheme. 

To understand the effect of the $\hat{x}$ strain field on the gate fidelity, we begin by plotting its effect on the non-zero state-preserving and state-flipping transition amplitudes at the ``magic'' frequency [see Fig.~\ref{fig:strain_field}(b)]. We observe that in the presence of a small $\hat{x}$ strain field the state-flipping transitions are only slightly affected, while the state preserving transition amplitudes are still suppressed. 

In scheme $M1$, state-preserving photons can be blocked by the polarizer on the collection path due to the orthogonality property of the dipole moment matrix elements. As the state-flipping transitions are only slightly affected by the $\hat{x}$ direction strain field, the gate fidelity of $M1$ is almost flat [see Fig.~\ref{fig:strain_field}(a)]. 

When the drive is polarized along $(\hat{x}+\hat{y})$ direction, since the transition amplitudes $A_{f,2}^{x}$ and $A_{f,3}^{x}$ only slightly affected by the $\hat{x}$ strain field [see Fig.~\ref{fig:strain_field}(b)], neither the rates nor the polarizations of the state-flipping photons are heavily affected [see green and red curves in Fig.~\ref{fig:strain_field}(d)]. However, the $\hat{x}$ strain field shifts the four state preserving transition amplitudes a lot, which causes the increase of the state-preserving transition rates [see blue and orange curves in Fig.~\ref{fig:strain_field}(d)]. Note, the fact that the polarization of the state-preserving photons points along $(\hat{x}-\hat{y})$ direction without strain field is because the state-preserving transition amplitudes satisfy $A_{p,2}^{x}=A_{p,3}^{y} = - A_{p,2}^{y} = - A_{p,3}^{x}$ at the ``magic'' point. The non-zero $\hat{x}$ strain field destroys this feature, which causes the polarization of the state-preserving photons deviates from $(\hat{x}-\hat{y})$ direction [see Fig.~\ref{fig:strain_field}(d) top panel].

In $M2$, the polarizer on the collection path is along $(\hat{x}-\hat{y})$ direction, which still allows most of the state-flipping photons passing through. In non-perturbed case, the state-preserving photons are polarized along $(\hat{x}-\hat{y})$ direction, which can pass the collection path polarizer for certain. With the $\hat{x}$ strain field perturbation, the more the polarization of the state-preserving photons deviates from $(\hat{x}-\hat{y})$ direction, the less probable the photon can pass the collection path polarizer. However, the $x$-direction strain field also boost the generation rates of the state-preserving photons [see Fig.~\ref{fig:strain_field}(d) bottom panel]. Combining these two factors, the overall gate fidelity for scheme $M2$ drops to $\sim 0.95$ as the $x$-direction strain field increases to $1$~GHz [see Fig.~\ref{fig:strain_field}(a)].

However, in scheme $M3$, the collection path $(\hat{x}+\hat{y})$ polarizer blocks most of the state-flipping photons, which makes this scheme fragile to the leaking state-preserving photons. The key for the success of $M3$ in the non-perturbed case is the fact that state-preserving photons is polarized along $(\hat{x}-\hat{y})$ direction. However, as we increase the $\hat{x}$ strain field, the polarization of the state-preserving photons are not exactly aligned $(\hat{x}-\hat{y})$ direction [see Fig.~\ref{fig:strain_field}(d) top panel], which deteriorates the gate fidelity as shown in Fig.~\ref{fig:strain_field}(a).

The entanglement gate fidelity $F_e$ when $\hat{y}$ direction strain field is applied to the diamond crystal is plotted in Fig.~\ref{fig:strain_field}(b). The $\hat{y}$ direction strain field mixes the states $\ket{e_1} \leftrightarrow \ket{e_5}$, $\ket{e_2}\leftrightarrow\ket{e_6}$, and $\ket{e_3}\leftrightarrow\ket{e_4}$. The mixing of the states results in the loss of the dipole moment orthogonality property. Therefore, for drive photons polarized along $\hat{x}$ direction the state-preserving photons are not necessarily polarized along $\hat{x}$, nor the state-flipping photons along $\hat{y}$. The polarization of both state-preserving photons and state-flipping photons relative to the $\hat{y}$ direction is plotted in Fig.~\ref{fig:strain_field}(c). As we vary the $\hat{y}$ direction strain field, the polarization of the two kinds of state-flipping photons remains nearly along the $\hat{y}$ direction, but the polarization of state-preserving photons changes significantly. 

For scheme $M1$ (with $\hat{x}$ polarized drive), there are two main sources of error: (1) unbalanced state-flipping transitions as before and (2) $\hat{y}$ photons from state-preserving transitions that leak past the polarizer. We plot the polarization angule with respect to the $\hat{y}$ direction and the magnitude of the transition amplitudes for both state-preserving and state-flipping transitions in Fig.~\ref{fig:strain_field}(e). As we increase the perturbation of $y$-direction strain field, the state-preserving transition amplitudes are slightly increased. Combining with the fact that polarization of the state-preserving photons are no longer along $\hat{x}$ direction exactly, the leaking state-preserving photons to the detector decreases the gate fidelity to $\sim 0.95$ as we change $\hat{y}$ direction strain field to $\sim \pm 1$~GHz. 

Similarly, when the drive is along $(\hat{x} + \hat{y})$ direction, the polarization features that were utilized in gate operation schemes $M2$ and $M3$ are no longer valid. We plot the deviation of the polarization angle of all scattered photons with respect to the polarizer direction in $M2$, i.e. $(\hat{x} - \hat{y})$, in the top panel of Fig.~\ref{fig:strain_field}(f). The polarization of the state-flipping photons are slightly affect by the $\hat{y}$-direction strain field, while the state-preserving photon polarization rotates $\sim 54^{\circ}$ as we increase $\hat{y}$-direction strain field to $\pm 1$~GHz. The amplitudes of the state-preserving and state-flipping transitions are plotted in the bottom panel of Fig.~\ref{fig:strain_field}(f). We observe that the transition amplitudes are only slightly affected by the applied $\hat{y}$-direction strain field. Therefore, to understand the effect of $y$-direction strain field on schemes $M2$ and $M3$, we mainly focus on the rotation of the scattered photon polarizations. 

The main error source in scheme $M2$ without perturbation is the leakage of  state-preserving photons past the polarizer in the collection path. As we change the $\hat{y}$-direction strain field, the state-preserving transitions are only slightly affected, while the polarization of the state-preserving photons rotates away from the collection path polarizer direction, i.e. $(\hat{x}-\hat{y})$ direction [see Fig.~\ref{fig:strain_field}(h)]. The state-preserving photons thus have a smaller probability to get past the polarizer in the collection path. Consequently, the gate fidelity for scheme $M2$ slightly improves as a result of $\hat{y}$-direction strain field perturbation, as we show in Fig.~\ref{fig:strain_field}(b).

On the other hand, the perfect gate fidelity of scheme $M3$ in the absence of perturbation is based on the fact that all state-preserving photons are polarized along $(\hat{x}-\hat{y})$ direction and hence are stopped by the polarizer in the collection path (along with most of the state-flipping photons). Large rotation angle of the state-preserving photon polarization makes the leakage rate of the state-preserving photons comparable to that of the state-flipping photons. This quickly degrades the gate fidelity as we show in  Fig.~\ref{fig:strain_field}(b).

\bibliography{NV_Entanglement}

\end{document}